\documentclass{article}

\usepackage{PRIMEarxiv}

\usepackage[utf8]{inputenc} % allow utf-8 input
\usepackage[T1]{fontenc}    % use 8-bit T1 fonts
\usepackage{url}            % simple URL typesetting
\usepackage{booktabs}       % professional-quality tables
\usepackage{amsfonts}       % blackboard math symbols
\usepackage{nicefrac}       % compact symbols for 1/2, etc.
\usepackage{microtype}      % microtypography
\usepackage{lipsum}
\usepackage{fancyhdr}       % header
\usepackage{graphicx}       % graphics
\graphicspath{{media/}}     % organize your images and other figures under media/ folder
\usepackage{amsmath}
\usepackage{bm}
\usepackage{listings}
\usepackage{makecell}
\usepackage{float}
\usepackage{anyfontsize}
\DeclareUnicodeCharacter{2113}{l}
\title{Physics-informed AI and ML-based sparse system identification algorithm for discovery of PDE's representing nonlinear dynamic systems
}

\author{
  Ashish Pal, Sutanu Bhowmick, Satish Nagarajaiah \\
  Department of Civil and Environmental Engineering \\
  Rice University \\
  Houston, Texas, USA, 77005\\
  \texttt{satish.nagarajaiah@rice.edu} \\
}

\begin{document}
\maketitle

\begin{abstract}
Sparse system identification of nonlinear dynamic systems is still challenging, especially for stiff and high-order differential equations for noisy measurement data. The use of highly correlated functions makes distinguishing between true and false functions difficult, which limits the choice of functions. In this study, an equation discovery method has been proposed to tackle these problems. The key elements include a) use of B-splines for data fitting to get analytical derivatives superior to numerical derivatives, b) sequentially regularized derivatives for denoising (SRDD) algorithm, highly effective in removing noise from signal without system information loss, c) uncorrelated component analysis (UCA) algorithm that identifies and eliminates highly correlated functions while retaining the true functions, and d) physics-informed spline fitting  (PISF) where the spline fitting is updated gradually while satisfying the governing equation with a dictionary of candidate functions to converge to the correct equation sequentially.  The complete framework is built on a unified deep-learning architecture that eases the optimization process. The proposed method is demonstrated to discover various differential equations at various noise levels, including three-dimensional, fourth-order, and stiff equations. The parameter estimation converges accurately to the true values with a small coefficient of variation, suggesting robustness to the noise. 
\end{abstract}

% keywords can be removed

\section{Introduction}
The guiding physical principles of dynamic systems found across different fields of science and engineering are described with various types of nonlinearity. Nonlinear dynamics are not just limited to engineering structures such as buildings\cite{lai2019semi}, bridges \cite{bhowmick2023physics}, and wind turbines\cite{jimenez2020maintenance}, but systems in biological/biomedical sciences \cite{alber2019integrating}, chemistry\cite{sharma2022hybrid}, and control protocols \cite{wang2021neural} are also modeled as nonlinear dynamic systems, which makes the problem of nonlinear system identification relevant to a broader community.

The identification of the correct type of nonlinearity has been of keen interest to researchers, which is achieved either from first principles or using the measured data to solve the inverse problem (data-driven methods). With the advancement in computational power, machine-learning algorithms, and sensors to collect high-dimensional data, data-driven methods seem to be more viable. These methods include Koopman operator \cite{bruder2019nonlinear}, dynamic mode decomposition \cite{klus2020data}, subspace methods \cite{marchesiello2008time},  machine learning methods based on artificial neural networks \cite{ayala2020nonlinear,yu2019system, li2020fourier} and deep learning \cite{silvestrini2022deep,wu2019deep,de2016randomized}, physics-informed machine learning \cite{raissi2018hidden,bhowmick2023data,raissi2019physics,lai2021structural}, Kalman-filter based methods \cite{pal2024sparsity,zhang2017structural}, symbolic regression \cite{bomarito2021development,reinbold2021robust,sun2019data}, and sparse system-identification methods \cite{lai2019sparse, cortiella2021sparse, rudy2017data}. For a full understanding of a nonlinear system, the employed method must recover a mathematical model representing the physical principles of the system and not just relate the input and output by creating a black box. The most popular method that recovers the governing equation of the nonlinear dynamic systems conveniently and computationally efficiently is the sparse system-identification method.

In the sparse system-identification method, a dictionary of linear/nonlinear functions is created, and they are linearly superimposed to represent the governing equation of the system's dynamics. Typically, the number of functions comprising the governing equation is few compared to the total number of functions in the library. Therefore, the solution to this problem is obtained through enforcing sparsity. The major obstacles in this formulation are i) noise in the measurement data, ii) sensitivity to the threshold value used for sparse regression, and iii) the presence of highly correlated functions in the library that cannot be eliminated using thresholding.

First, due to the noise in measurement data, the signal derivatives calculated via numerical differentiation are even noisier, which poses a challenge in the identification of higher-order ordinary/partial differential equations (ODE/PDE) and stiff-differential equations \cite{rudy2017data}. Several alternatives to numerical differentiation have been proposed, which include polynomial interpolation \cite{rudy2017data}, spectral method \cite{schaeffer2017learning}, automatic differentiation \cite{long2018pde,long2019pde,both2021deepmod,xu2021robust,raviprakash2022hybrid,stephany2022pde}, data approximation using neural networks followed by numerical differentiation \cite{rao2022discovering,zhang2022parsimony}, and data approximation using B-splines followed by analytical differentiation \cite{bhowmick2023physics,bhowmick2023data}. The analytical differentiation of B-splines looks the most promising as the derivative calculation is exact, and it has proved to be robust to 100\% noise in the measured data and works for stiff-differential equations \cite{bhowmick2023data}.

Second, during the sparse regression, the threshold value (below which functions are eliminated from the candidacy) must be tuned to the optimal value for correct identification \cite{raviprakash2022hybrid}. Across studies in the existing literature, sparse estimation has been achieved using strategies such as L\textsubscript{1} regularization \cite{both2021deepmod,long2019pde}, sequentially thresholded least squares \cite{rudy2017data}, and sequentially thresholded ridge regression \cite{raviprakash2022hybrid}. These methods converge to the correct solution under the condition that the coefficients of false functions (functions not part of the governing equation) remain small. This condition does not hold for several scenarios, such as in the presence of higher noise levels in the measured data (especially for stiff-differential equations) and the inclusion of certain functions in the library that highly correlate with the true functions (functions part of the governing equation). In these situations, false functions survive the sparse regression, and sometimes their coefficients are higher than those of the true functions \cite{pal2024sparsity}, making sparse regression a sole strategy ineffective.

In this study, we propose an algorithm for sparse system identification via physics-informed deep learning to tackle all the above-mentioned problems. There is better flexibility in including terms in the library that are problematic for sparse system identification in the existing algorithms. The proposed algorithm takes the measured noisy data as the input and provides the governing equation of the multi-dimensional process with the correct terms and accurate parameter values. The algorithm consists of four major components and brings novelty in the following ways:
\begin{enumerate}
    \item Perform analytical differentiation via B-spline functions for accuracy. A novel sequentially L\textsubscript{1}-regularized derivatives for denoising (SRDD) algorithm that can generate accurate spatial/temporal derivatives to any order.
    \item Sequentially thresholded regression with L\textsubscript{1} regularization to speed up the sparse regression step using deep learning.
    \item An algorithm for uncorrelated component analysis (UCA) based on singular value decomposition (SVD) and cross-correlation (CC) to remove highly correlated functions that survive the sparse regression.
    \item Physics-informed spline fitting (PISF) based sequential elimination of any remaining false functions for robust system identification.
    \item A single deep-learning architecture that can perform the above tasks individually or simultaneously based on the loss function.
\end{enumerate}
In the present literature, the sparse system identification methods focus on two tasks: (i) denoising the measured data to get accurate derivatives and (ii) algorithms to find the correct sparse estimate for equation discovery. These two tasks are performed independently using the existing methods. However, in the presented study, the element of PISF is introduced, which performs data fitting and equation discovery simultaneously. This is a physics-informed method of fitting the measured data to follow the basic underlying physics that paves the way for a robust method for equation discovery. The coefficients of the B-spline basis curves are updated so that the fitted curve simultaneously represents the measured data and the system's governing equation. PISF accurately discovers the correct equation and estimates the parameters even for higher-order ODE/PDE, stiff-differential equations, and high noise levels.

\section{Physics informed deep learning framework for sparse identification of nonlinear dynamic systems}

\begin{figure}[h]
\centering
\includegraphics[width=\textwidth]{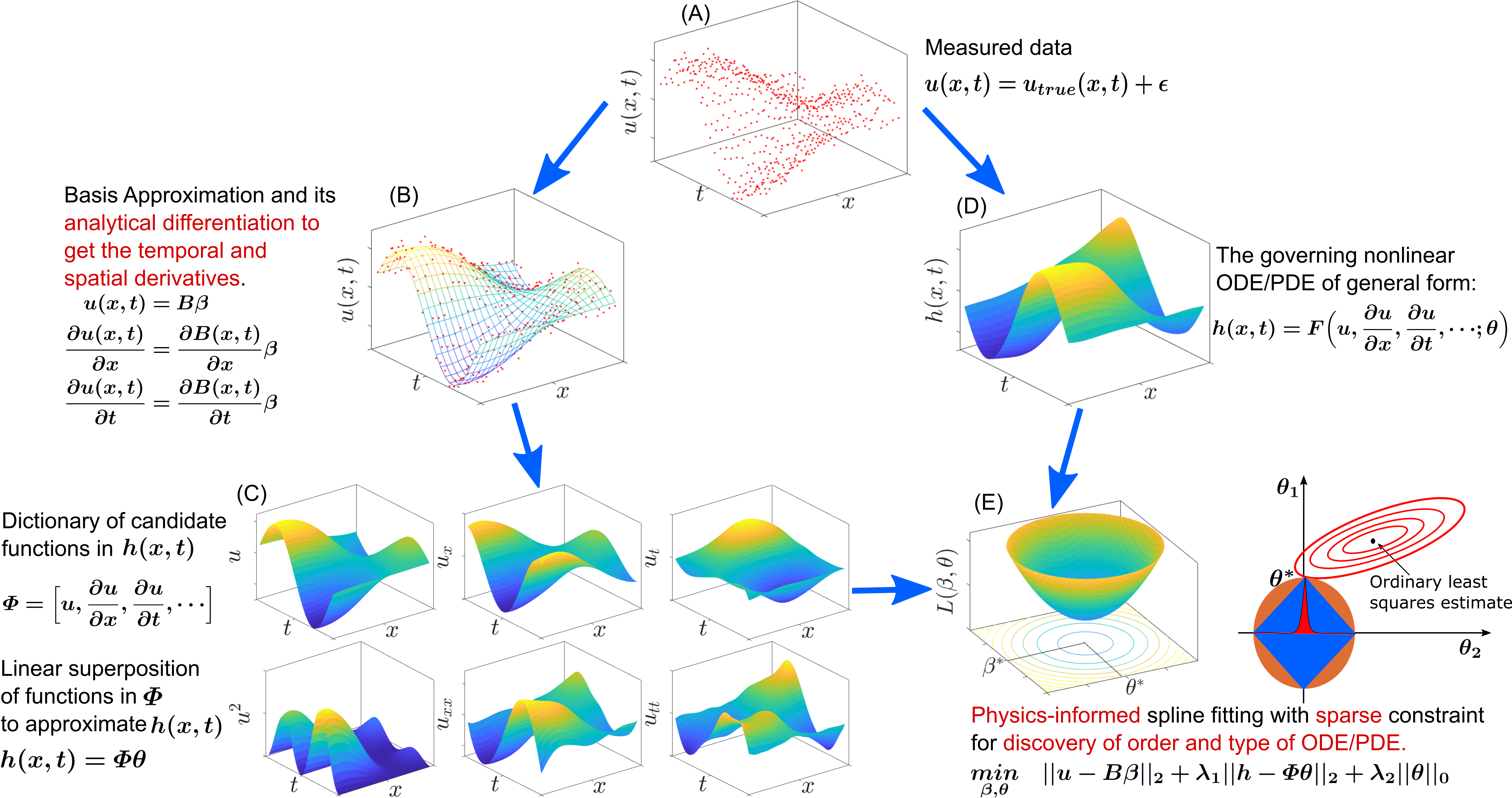}
\caption{Proposed framework for equation discovery of nonlinear dynamic systems. (A) The measured noisy field data from a general two-dimensional dynamic system at discrete points (B) Approximation of the measured data using the linear combination of B-spline basis functions with unknown coefficients $\bm{\beta}$. (C) Generation of the library comprising of various terms via the fitted B-spline functions. The spatial and temporal derivatives are analytically derived using the B-spline functions. (D) The underlying governing equation $h(x,t)$ with unknown parameters $\bm{\theta}$ the dynamic system follows, which needs to be discovered. (E) The global optimization function for equation discovery requires the simultaneous approximation of the measured data with B-spline functions to obtain $\bm{\beta^*}$, satisfying the governing equation utilizing the dictionary of candidate functions, and sparse estimation of the corresponding parameters $\bm{\theta^*}$. }\label{fig:main_idea}
\end{figure}

\begin{figure}[h]
\centering
\includegraphics[width=\textwidth]{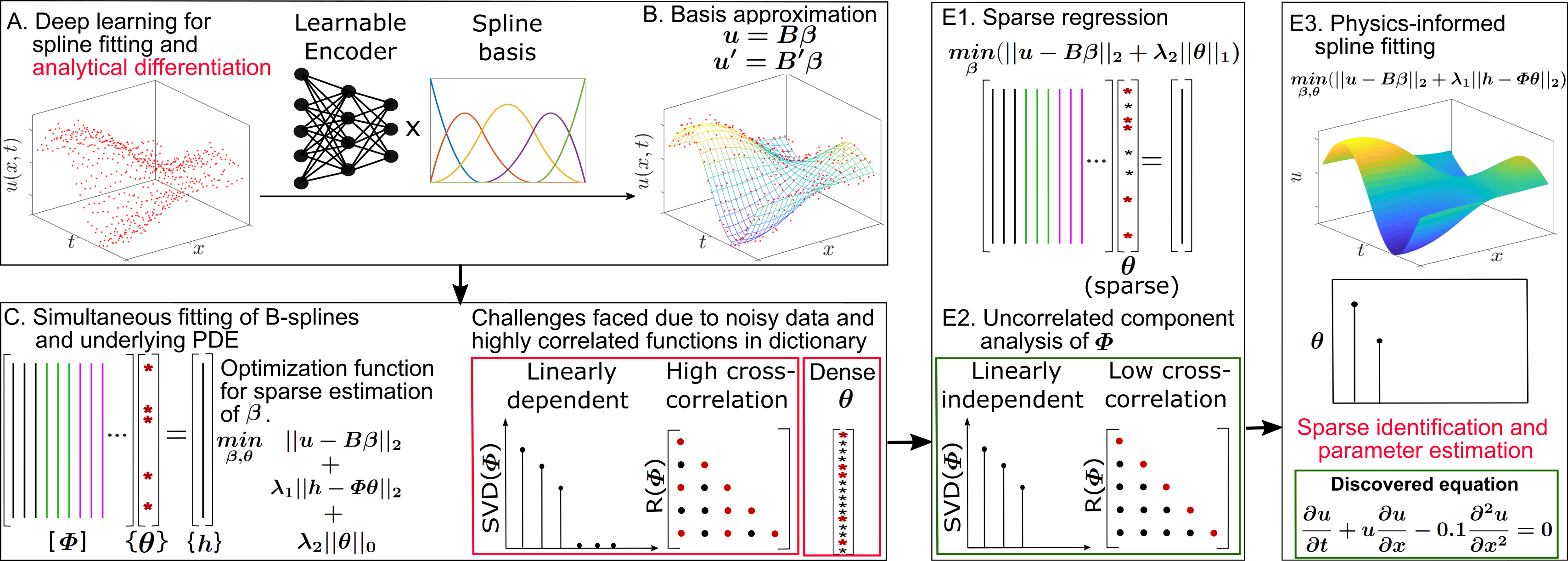}
\caption{A brief description of challenges in sparse system identification and proposed solution strategy. (A) The measured noise data is input to the deep learning network that passes through an encoder and provides scaling factors $\bm{\beta}$ for B-spline fitting (B) The approximation of measured data using B-spline functions and $\bm{\beta}$ obtained from A (C) The setup of the regression formulation using the chosen library of functions for equation discovery. The noise and highly correlated functions pose a challenge to sparse system identification (E1) Sparse regression using L\textsubscript{1}-regularization to remove small contribution functions (E2) Uncorrelated Component Analysis to remove highly correlated functions that provide repetitive information (E3) Physics-informed spline fitting for simultaneous data fitting and equation discovery.}\label{fig:main_idea2}
\end{figure}

The framework of the proposed methodology is shown in Figure \ref{fig:main_idea}. The physics of a system is often modeled as a differential equation. The measured response data (Figure \ref{fig:main_idea}A) of the system follows the same governing equation and, hence, can be utilized to identify it. The measured data is approximated by fitting B-spline functions defined over the complete domain to get analytical forms of the derivatives, thus avoiding the numerical differentiation (Figure \ref{fig:main_idea}B). The basis functions are derived, and each of them is appropriately scaled before linearly superimposing to fit the measured data as shown below:
\begin{equation}
    \bm{u}(x,t)=\bm{B}(x,t)\bm{\beta}
\end{equation}
where $\bm{B}$ are the B-spline functions, and $\bm{\beta}$ are the spline scaling factors (further details in the Appendix). A nonlinear dynamic system can be represented as follows:
\begin{equation}
    \frac{\partial^n \bm{u}(x,t)}{\partial t^n} = \bm{h}(u(x,t))
\end{equation}
where, $u(x,t)$ is the $L \times 1$ vector of measured response field in time $t$ and space $x$, and $\bm{h}()$ is the governing differential equation (Figure \ref{fig:main_idea}D). This study aims to find $\bm{h}()$. Since the form of $\bm{h}()$ is unknown, it is substituted with a linear combination of a library of functions ($\bm{\Phi}$) that spans the possible terms in $\bm{h}()$ (Figure \ref{fig:main_idea}C). $\bm{\Phi}$ can consist of polynomial functions, partial derivatives of single variables as well as mixed variables, cross-multiplication of polynomial functions with partial derivative functions, etc. An example library is shown below:
\begin{equation}
    \bm{\Phi}(\bm{u}) = 
    \begin{bmatrix}
        \bm{u} & \bm{u}^2 & \cdots & \frac{\partial \bm{u}}{\partial t} & \frac{\partial \bm{u}}{\partial x} & \frac{\partial^2 \bm{u}}{\partial t^2}  & \frac{\partial^2 \bm{u}}{\partial x^2} &\frac{\partial^2 \bm{u}}{\partial t \partial x} & \cdots
        & u\frac{\partial \bm{u}}{\partial t} & u\frac{\partial \bm{u}}{\partial x} & \cdots
    \end{bmatrix}_{L \times m}
\end{equation}
The differential equation $\bm{h}()$ in equation 2 is replaced with a linear combination of functions in $\bm{\Phi}$ as follows:
\begin{equation}
\begin{split}
    \frac{\partial^n \bm{u}(x,t)}{\partial t^n} = &\bm{\Phi}(\bm{u})\bm{\theta}\\
    h(x,t) = &\bm{\Phi}(\bm{u})\bm{\theta}
\end{split}
\end{equation}
where $\bm{\theta}$ is $m \times 1$ vector of coefficients of $\bm{\Phi}$, $m$ is the number of functions in $\bm{\Phi}$. Under the assumption that the number of terms in $\bm{h}()$ is few, the solution to equation 4 corresponds to the sparse estimate of $\bm{\theta}$. The objective of this study is to find the spline fitting of the measured data that follows the system's governing equation by solving the following optimization problem combining equations 1 and 4 (Figure \ref{fig:main_idea}E):
\begin{equation}
    \mathrm{min}_{\bm{\beta},\bm{\theta}} \quad ||\bm{u}-\bm{B\beta}||_2 + \lambda_1||h-\bm{\Phi\theta}||_2+\lambda_2||\bm{\theta}||_0
\end{equation}
The loss terms $||\bm{u}-\bm{B\beta}||_2$, $||h-\bm{\Phi\theta}||_2$ are the error in the fitting of the B-splines to the measured data and the governing equation, respectively, and $||\bm{\theta}||_0$ applies a sparse regularization on the coefficient vector $\bm{\theta}$. $\lambda_1$ is the hyperparameter that controls the relative contribution of equation fitting with data fitting and $\lambda_2$ controls the strength of regularization on $\bm{\theta}$.\\
The optimization problem shown in equation 5 is non-convex and cannot be solved in a single step. Figure \ref{fig:main_idea2} illustrates the challenges in sparse system identification and a three-step solution to tackle them. Due to the noise in the measured data and the high correlation among functions, the $\bm{\theta}$ estimate from regression is dense (see Figure \ref{fig:main_idea2}C). The SVD of the library shows several singular values with tiny contributions suggesting repetitive information, and the CC matrix indicates the functions that show high correlation. In the proposed methodology, first, the sparse regression is used to eliminate all the functions that show little contribution (see Figure \ref{fig:main_idea2}E1). Second, UCA is performed to eliminate correlated functions that are not part of the governing equation but still show high contribution (see Figure \ref{fig:main_idea2}E2). Third, physics-informed spline fitting is used to sequentially eliminate the lowest contributing functions till the convergence criteria are met (see Figure \ref{fig:main_idea2}E3). Before discussing each of these steps in detail, the deep learning network, B-spline functions, and data-denoising algorithm must be introduced briefly.

\subsection{Deep learning architecture}
\begin{figure}[H]
\centering
\includegraphics[width=0.8\textwidth]{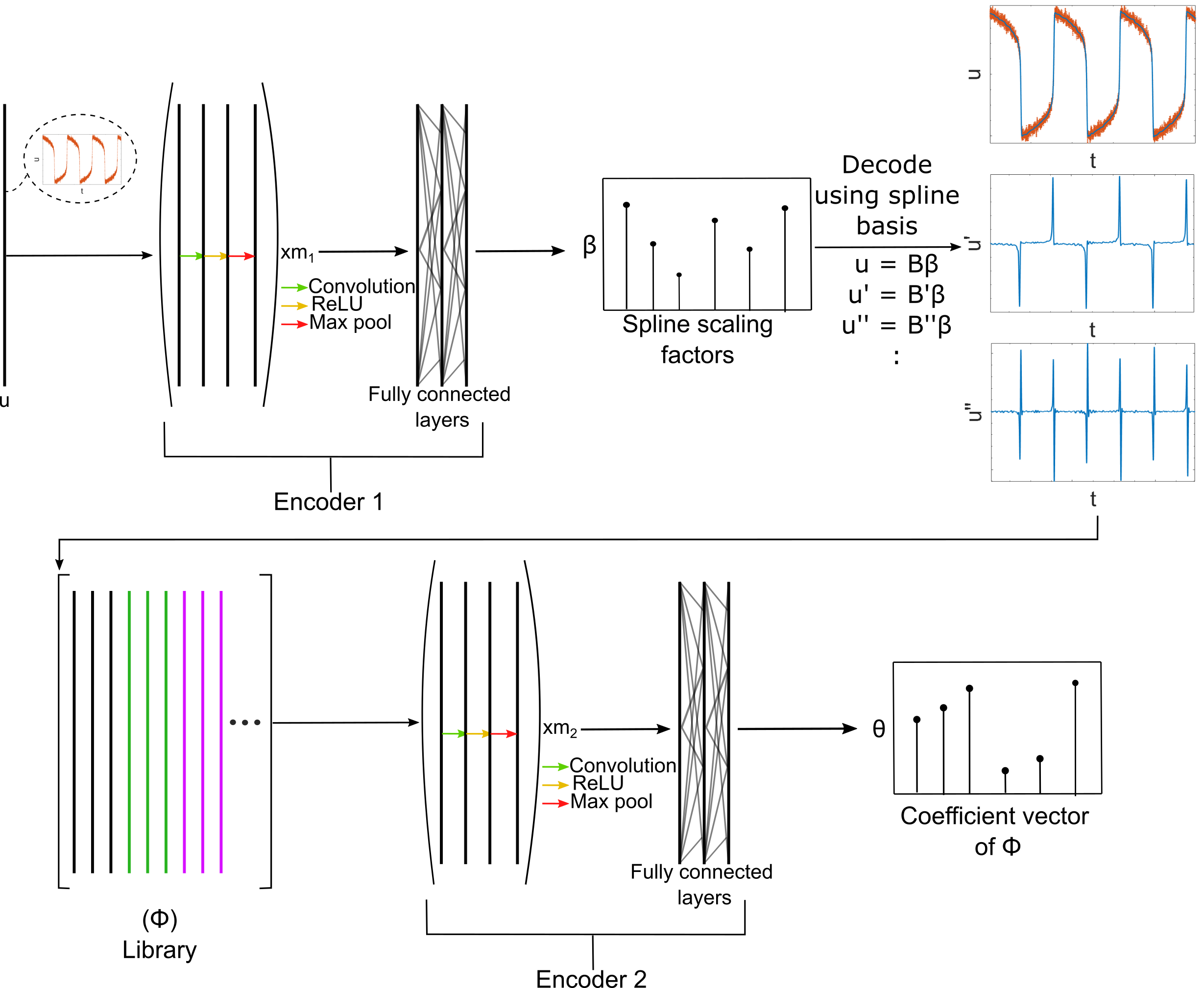}
\caption{Deep learning architecture}
\label{fig:architecture}
\end{figure}

The task of spline fitting (estimating $\bm{\beta}$) and equation fitting (estimating $\bm{\theta}$) is achieved using deep learning. The network has the architecture as shown in Figure \ref{fig:architecture}. The architecture includes the input $u$, Encoder 1, spline scaling factors $\bm{\beta}$, Encoder 2, and coefficient vector $\theta$. Encoder 1 and Encoder 2 comprise the trainable parts of the architecture, where, Encoder 1 is responsible for estimating $\bm{\beta}$ and Encoder 2 for estimating $\bm{\theta}$. Encoder 1 consists of $m_1$ sets of convolution layer + REctifiled Linear Unit (RELU) + max pool layer, followed by two fully connected layers. The parameters of the convolution layers such as the filter size, number of filters, and the value of $m_1$ are decided based on the size of the input $u$. The size of the last fully connected layer is equal to the length of $\bm{\beta}$. The design of Encoder 2 is similar to Encoder 1 with changes in the filter size, number of filters, and value of $m_2$ based on the size of the dictionary $\bm{\Phi}$. The size of the last fully connected layer is equal to the length of $\bm{\theta}$. Convolution neural network has been chosen because the convolution operation captures the spatial/temporal correlations and features that are expected to provide better estimates of the spline basis coefficients $\bm{\beta}$ while data fitting. Similarly, functions in the dictionary are also series in time/space which show correlation with the neighboring values.

\subsubsection{Loss functions and hyperparameters}
Four types of losses are defined for this network that are used in different combinations for carrying out different tasks in the proposed framework. The total loss of the network is defined as follows:
\begin{equation}
    loss = c_1u_{loss} + (c_{2t}u_{loss}^{t}+c_{2x}u_{loss}^{x}) + c_3\theta_{loss} + c_4P_{loss}\\
\end{equation}
\begin{equation}
    u_{loss} = ||\bm{u}-\bm{B\beta}||_2
\end{equation}
\begin{equation}
    u_{loss}^{t}, u_{loss}^{x}= \Big|\Big|\frac{\partial^{i} \bm{u}}{\partial t^{i}}\Big|\Big|_1, \Big|\Big|\frac{\partial^{j} \bm{u}}{\partial x^{j}}\Big|\Big|_1
\end{equation}
\begin{equation}
    \bm{\theta}_{loss} = ||w\bm{\theta}||_1
\end{equation}
\begin{equation}
    \bm{P}_{loss} = ||h-\bm{\Phi\theta}||_2
\end{equation}
where, $u_{loss}$ is the mean-squared error (MSE) in the spline fitting, $u_{loss}^{t}, u_{loss}^{x}$ are the L\textsubscript{1} norm of the $i^{th}$ time derivative and $j^{th}$ space derivative, respectively, $\theta_{loss}$ adds weighted L\textsubscript{1} penalty to the $\bm{\theta_{loss}}$ with a weight vector $\bm{w}$, and $P_{loss}$ is the MSE in the governing equation fitting. $c_1$, $c_{2t}$, $c_{2x}$, $c_3$, and $c_4$ are the hyperparameters controlling the contribution of different losses to the total loss. The vector $\bm{w}$ is proportional to the root mean square value (RMS) of the corresponding function in $\bm{\Phi}$ and the function complexity. More complex functions such as higher-order derivatives and compound functions will be assigned a higher weight. Scaling the coefficient with the RMS value of the function indicates the amount it contributes to the governing equation which is a more reasonable metric to eliminate functions. Implementation of each component of the proposed framework using the deep learning network is explained in detail in the following sections.

\subsection{B-splines}
B-spline functions are polynomial functions defined by order $o$ and $k$ number of knots spread over the domain with end knots positioned at the domain boundary. The total number of basis functions $m$ is equal to ($k-0+1$). A B-spline function of any degree can be obtained from lower-degree B-spline functions using the recursive Cox-de-Boor algorithm \cite{de1978practical}. An $o$-th order B spline function is ($o-1$) times continuously differentiable for non-repeated knots. The B-spline function can be analytically differentiated which provides smooth derivatives. A single B-spline function although defined over the full domain is nonzero for a small region and has compact support. A basis of B-spline functions can be used to generate smooth curves and its analytical derivatives over the complete domain using linear superposition as described below:
\begin{equation}
\begin{split}
    f(x) &= B(x)\beta \\
    \frac{\partial^n f(x)}{\partial x^n} &= \frac{\partial^n B(x)}{\partial x^n}\beta \quad \forall \quad n=1,2,\cdots,o-1
\end{split}
\end{equation}
where, $f$ is a $N \times 1$ vector defined over the domain $x$, $B$ is $N \times m$ matrix formed of $m$ number of B-spline basis functions, and $\beta$ is $m \times 1$ vector of scaling factors for the corresponding basis functions. The extension of 1D B-spline basis to $n$D B-spline basis is straightforward. A single $n$D B-spline function defined over $n$ dimensions is formed by outer products of $n$ 1D B-spline functions. A function defined in $n$ dimensions can be represented as the linear combination of $n$D spline basis as described below:
\begin{equation}
    B^i(x_1,x_2,\cdots,x_n)_{\{N_1\times N_2\times\cdots\times N_n\}}=B_1^{i1}(x_1)_{\{N_1\times 1\}}\otimes B_2^{i2}(x_2)_{\{N_2\times 1\}}\otimes \cdots\otimes B_n^{in}(x_n)_{\{N_n\times 1\}}
\end{equation}
\begin{equation}
    f(x_1,x_2,\cdots,x_n) = \sum_{i1=1}^{m_1}\sum_{i2=1}^{m_2}\cdots\sum_{in=1}^{m_n}B^i(x_1,x_2,\cdots,x_n)\beta_i
\end{equation}
where $B^i$ is a nD B-spline basis function, $B_j$ is $N_j \times m_j$ matrix of 1D B-spline in the $j^{th}$ dimension. For further details, refer to \cite{de1978practical,bhowmick2023data}.

\subsection{SRDD method and analytical differentiation of the data}
\begin{figure}[h]
\centering
\includegraphics[width=0.5\textwidth]{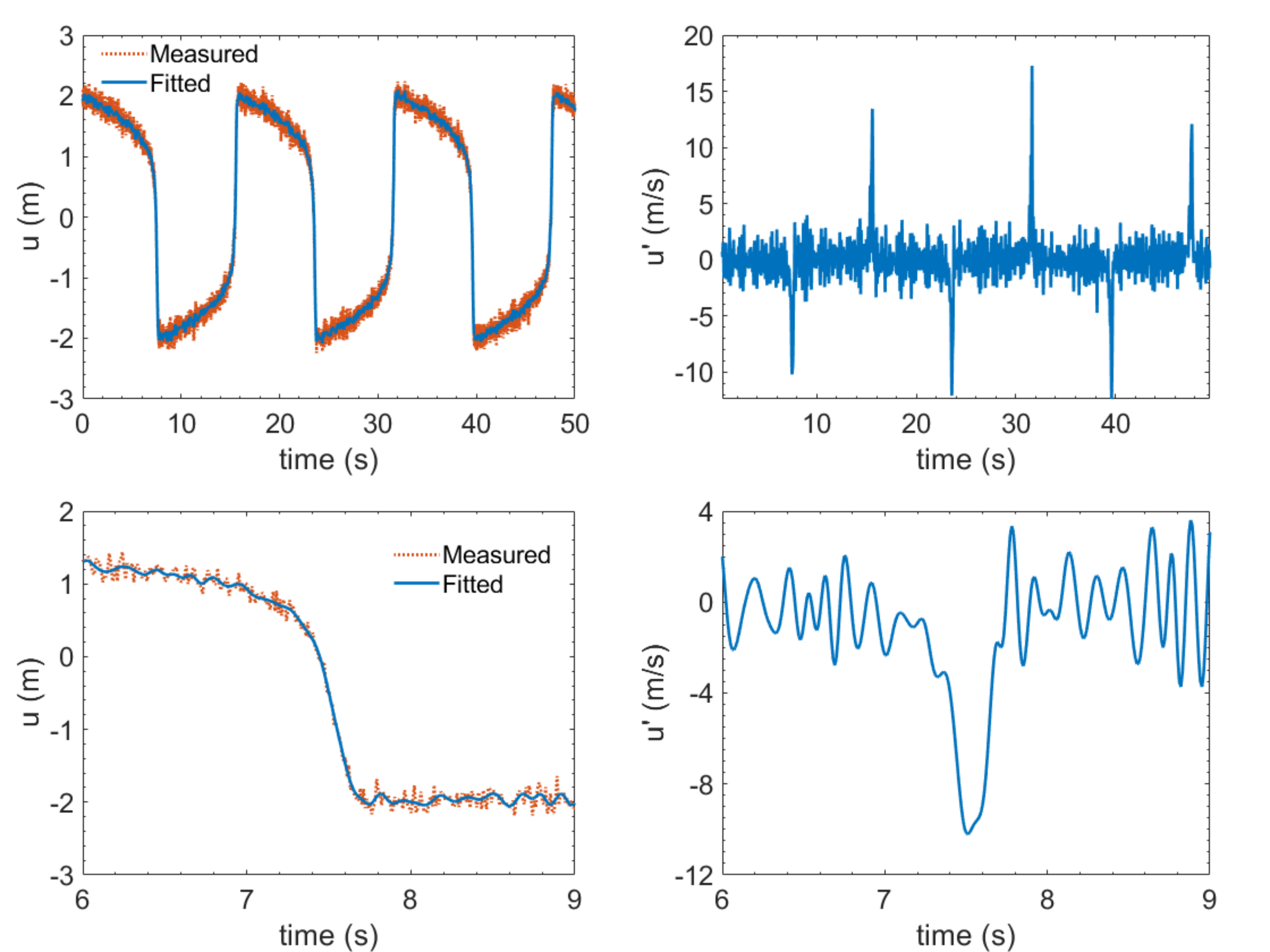}
\caption{Spline fitting to the measured data from Van der Pol oscillator, analytically derived first derivative and a closeup look from 6-9 seconds.}
\label{fig:vanderpol_fit}
\end{figure}

\begin{figure}[h]
\centering
\includegraphics[width=0.9\textwidth]{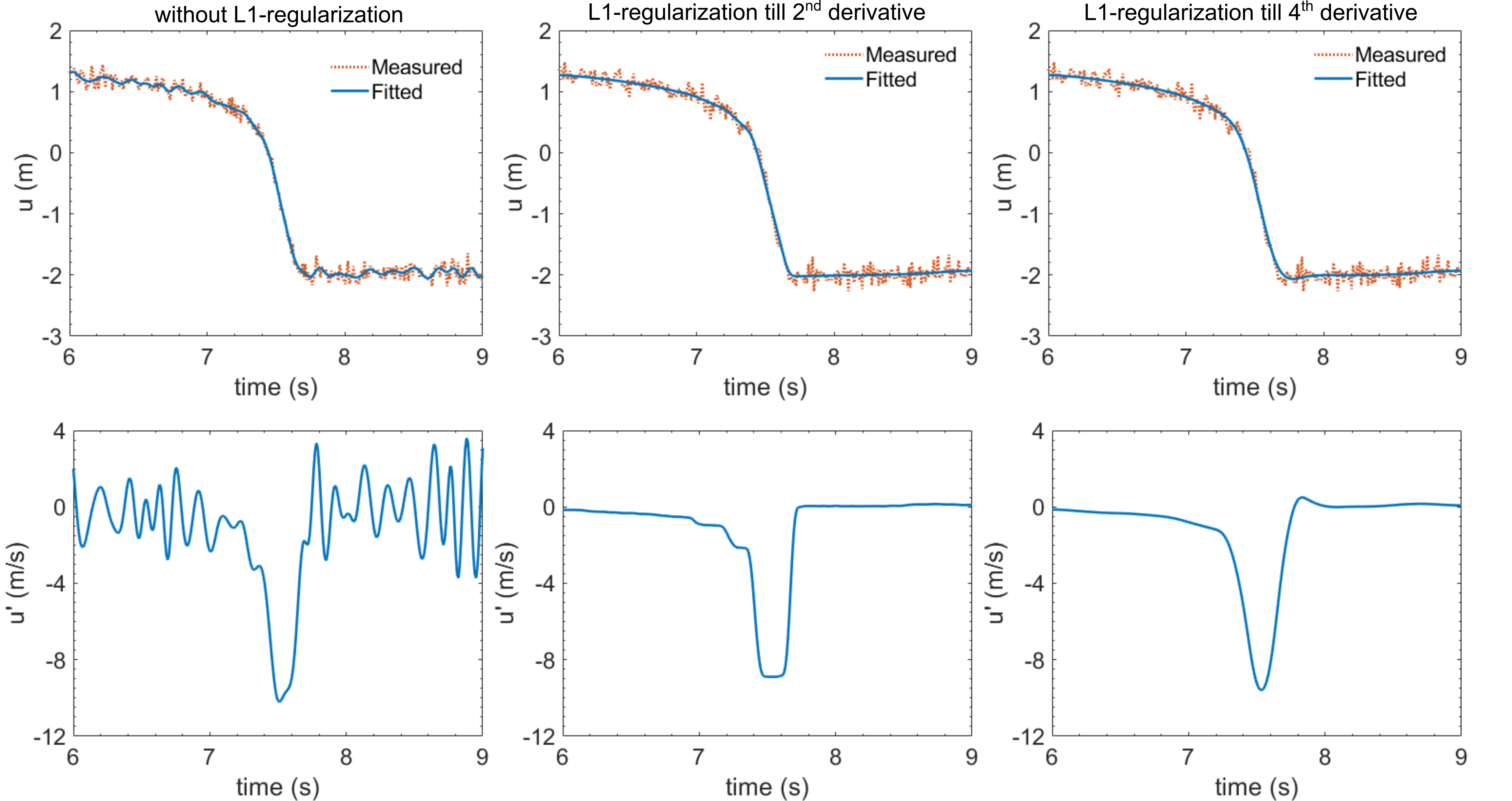}
\caption{Spline fitting and its first derivative with a) no regularization, b) regularization till 2\textsuperscript{nd} derivative, and c) regularization till 4\textsuperscript{th} derivative.}
\label{fig:vanderpol_deri}
\end{figure}

The noise in the measured data causes significant difficulties in the identification of partial differentiation equations. The numerical differentiation of the measured signal causes magnification of the noise and the effects increase with increasing derivative order. For stiff-differential equations, this effect gets even more magnified due to the rapid changes in the system states.\\
The B-spline basis curves are fitted to the measured data by minimizing $c_1u_{loss}$. To do so, the value of the constants in equation 6 is set as $c_1>0$, and $c_{2t}=c_{2x}=c_{3}=c_{4}=0$. This configuration of the constants trains Encoder 1 of the network which provides the values of $\bm{\beta}$. This $\bm{\beta}$ is substituted in equation 1 to get the fitted signal and its analytical derivatives. Figure \ref{fig:vanderpol_fit} shows the fitting for the displacement response of a Van der Pol oscillator with 10\% measurement noise (SNR=10). A closer look into a section of the signal from 6-9 seconds shows a smooth fit to the measured data, but, the first derivative shows higher levels of noise. The fitting is now updated by minimizing ($c_1u_{loss} + c_{2t}u_{loss}t + c_{2x}u_{loss}^x$). For $c1>0$, the value of the constants $c_{2t}$ and $c_{2x}$ are set such that, 
\begin{equation}
\begin{split}
    c_{2t}u_{loss}^{t}/c_1u_{loss} &=r_t\\
    c_{2x}u_{loss}^{x}/c_1u_{loss} &=r_x
\end{split}
\end{equation}
where $r_t$ and $r_x$ are the hyperparameters that control the relative importance of different losses. The value of $i,j$ in equation 8 must be chosen based on the amount of smoothing needed. The value of $i,j$ controls the smoothing of the fitted curve, regularizing lower-order derivatives smooths the curve more as compared to regularizing higher-order derivatives. To avoid over-smoothing, a lower limit is placed on the order of derivatives that will be L\textsubscript{1}-regularized, say $i_{min}, j_{min}$. L\textsubscript{1}-regularization works well in removing the unnecessary spikes but at the same time, it causes flattening of the peak and troughs in the signal, see Figure \ref{fig:vanderpol_deri}. Therefore, a sequential strategy is used to avoid this.

\subsection{Sparse regression}
\begin{figure}[h]
\centering
\includegraphics[width=\textwidth]{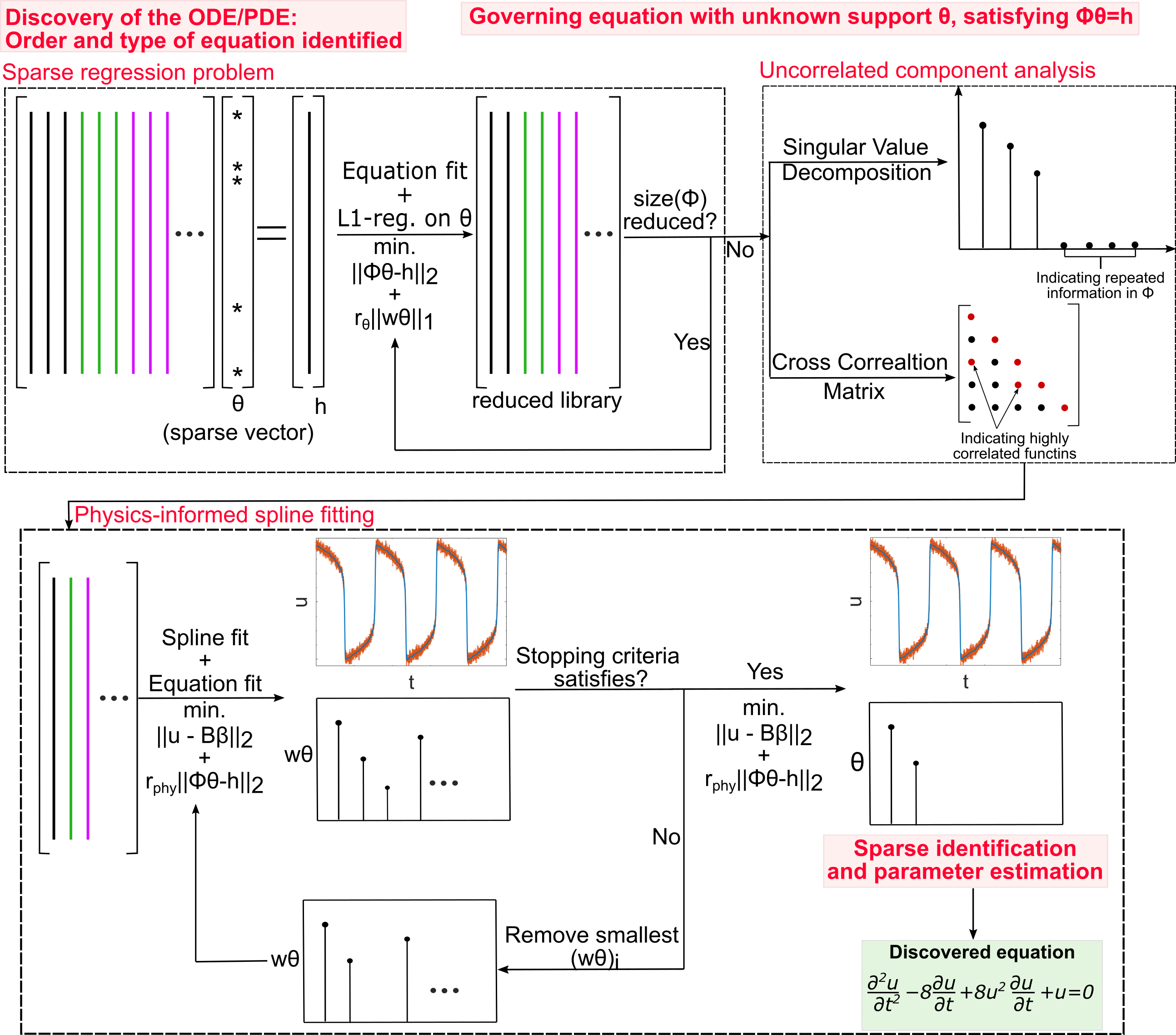}
\caption{Framework for the ODE/PDE discovery consisting of sparse regression, UCA, and PISF}
\label{fig:discover}
\end{figure}

As discussed in section 2, the identification process requires sparse estimation of $\bm{\theta}$ in equation 4. The sparse regression problem is solved by minimizing ($c_3\theta_{loss} + c_4P_{loss}$). For $c_4>0$, $c_3$ is set as follows:
\begin{equation}
    c_3\theta_{loss}/c_4P_{loss}=r_{\theta}
\end{equation}
where $r_{\theta}$ is the hyperparameter which decides how strongly sparsity is forced on the $\bm{\theta}$ vector. In most studies \cite{rudy2017data,long2019pde,raviprakash2022hybrid}, the sparsity is applied directly on the $\bm{\theta}$ vector, which are the coefficients of the functions in the library. In general, the value of the coefficient might not be the best metric to judge its importance/contribution. In the presence of noise, even false functions with small magnitudes start contributing by fitting the noise and showing up with bigger coefficient values. Therefore, a weight vector $\bm{w}$ is used for $\bm{\theta}$ which is proportional to the RMS value of the corresponding function. An additional factor included in the weight is the function complexity, which is defined by the number of operations performed to create that function. For example, the 3rd-order derivative is more complex than the 2nd-order derivative, a cubic polynomial is more complex than a quadratic polynomial, and a compound function is more complex than any of the participating functions.\\
The first section in Figure \ref{fig:discover} shows the iterative scheme used to get the sparse estimate of $\bm{\theta}$. This approach is similar to the sequentially thresholded least squares but with L\textsubscript{1}-regularization on $\bm{\theta}$ which accelerates the elimination process if using deep learning. The thresholding is applied to the contribution of the function ($\bm{\theta}_i\bm{\Phi}_i^{rms}$), not just on $\bm{\theta}_i$. The threshold value ($\Delta$) is set to 1\% of the maximum contribution from a single function. The solution obtained after minimizing ($c_3\theta_{loss} + c_4P_{loss}$) is thresholded for all $\bm{\theta}_i\bm{\Phi}_i^{rms}<\Delta$. If any function gets removed, then the above process is repeated with the remaining functions. The iterations are stopped when no further reduction is possible by thresholding.

\subsection{Uncorrelated Component Analysis}
Ideally, functions that are not in the governing differential equation should have zero contribution in the regression step in equation 4. For many functions, this is indeed the case, hence, they are identified and eliminated. However, due to noise in the measured data or sharing a high correlation with the true functions, some false functions show high contribution in the regression step due to which they can't be eliminated. If the cause is noise, then the pair of correlated functions show complementary behavior, i.e., these functions cancel each other's contribution. Otherwise, the false function is present due to a high correlation with a true function. In conclusion, if two functions share a high correlation, then either i) they are complementary functions, hence, both need to be eliminated or ii) one of them is a true function and the other is a false function.\\
In the second section of Figure \ref{fig:discover}, these highly correlated functions are removed by studying the Singular Value Decomposition (SVD) and cross-correlation (CC) of the remaining functions post sparse regression step. Functions that share a high correlation provide redundant information, which means columns of $\bm{\Phi}$ will lack independence. This information is easily found by finding the SVD of $\bm{\Phi}$. The number of singular values (say, $n_{rep}$) below a set threshold value, $S_{min}$, indicates the redundancy in the $\bm{\Phi}$ matrix. For $n_{rep}>1$, it is not possible to directly tell whether there are multiple sets of correlated functions or a group of correlated functions. Failing to identify this may result in the removal of true functions. Also, functions in $\bm{\Phi}$ share similar forms due to which it is common to have correlations among them. These are the major challenges addressed in the devised UCA algorithm to minimize the risk of removing true functions during the elimination process.\\

\subsection{Physics-informed spline fitting}
Due to noise in the measured data, a few functions can avoid elimination in the previous two steps. Till now, elimination has been carried out based on the initial fitting of the measured signal. In the last stage of the proposed framework, the deep learning network updates the spline fitting and simultaneously solves for $\bm{\theta}$ by minimizing ($c_1u_{loss} + c_4P_{loss}$). For $c_1>0$, $c_4$  is set as follows:
\begin{equation}
    c_4P_{loss}/c_1u_{loss}=r_{phy}
\end{equation}
where $r_{phy}$ is the hyperparameter that specifies the relative contribution of the signal fitting loss and the governing equation fitting loss. Doing this provides greater flexibility to the deep learning network to search for the combination of $\bm{\theta}$ that satisfies the governing equation by modifying the spline fitting itself. As shown in the third section of Figure \ref{fig:discover}, a sequential scheme is followed to remove any remaining false functions. At each iteration, spline fitting is adjusted to satisfy the governing equation with the remaining functions in $\bm{\Phi}$, and the function that has the minimum contribution is eliminated. The criteria to stop eliminations is defined based on the quality of spline fitting, the residual error in equation 4, SVD values of $\bm{\Phi}$, and the minimum contribution from a function. After the elimination of a function, refitting is performed, and $u_{loss}$ and $P_{loss}$ are calculated at each iteration. If $u_{loss}$ and $P_{loss}$ increase at any iteration above a set tolerance, then the previously eliminated function is put back into $\bm{\Phi}$. If the minimum SVD value of $\bm{\Phi}$ falls below a set threshold ($S_{min}$) or the minimum contribution of that function falls below $\Delta$ (definition of $S_{min}$ and $\Delta$ remains the same as in previous text), then it is removed and iterations continue, otherwise, we have the identified governing equation. The extra check for $S_{min}$ and $\Delta$ plays a critical role when measured data has high noise or in the case of stiff-differential equations. Noise in the measured signal gives the bandwidth for various combinations of functions to have similar $u_{loss}$ and $P_{loss}$. Having more functions in the dictionary makes it even easier to reduce $P_{loss}$ with a similar $u_{loss}$. Therefore, the additional check for $S_{min}$ and $\Delta$ helps to avoid the early stopping.

\section{Results}
The proposed method is applied to various ODE/PDE spanning stiff/non-stiff systems and fourth-order differential equations. The results of the equation discovery for the studied systems are shown in Table \ref{tab:results}. The subscripts 't,' 'x,' and 'y' denote derivatives with respect to time and two space dimensions, respectively.  Identification is performed for various noise levels, showcasing the robustness of the proposed method for high noise levels. The identification process is executed for 10 different bootstrap samples of noise-corrupted data at each noise level for repeatability.  For all the ODEs/PDEs, the parameter estimations are close to the true values. For the forced duffing oscillator and Burgers equation, the error is around 1\% even at 10\% noise level in the measured data. In the case of the Navier-stokes equation, which is a PDE in three dimensions with four terms to discover, the parameter estimation error is only 6\% for 10\% noise level. Similar performance is seen in the case of the Korteweg-De Vries equation and 2D wave equation. The highest error is observed in the case of the Kuramoto-Sivashinsky equation with 13\% error for 10\% noise level. Given the fourth-order PDE and the noise level, the parameter estimation is fairly accurate. The coefficient-of-variation (cov) indicates the robustness to noise with the highest cov value of 5.26\% in the Van der Pol oscillator case at 10\% noise level. Even for the fourth-order Kuramoto-Sivashinsky equation, the cov is at 1\% at 10\% noise level, which means the parameter estimation is consistent. Across all systems, cov is expectedly increasing with increasing noise level.\\
The parameter estimation is the final part of the proposed work. The critical part of equation discovery of the Van der Pol oscillator, Kuramoto-Sivashinsky equation, Navier-Stokes equation, and Korteweg-De Vries equation is discussed in detail in the following subsections.

\begin{table}[h]
\centering
\caption{Estimated parameters for different ODEs and PDEs for 1\% and 10\% noise in the measured data}
\resizebox{\textwidth}{!}{
\begin{tabular}{ccccc}%{|p{0.25\textwidth}|p{0.23\textwidth}|p{0.23\textwidth}|p{0.23\textwidth}|p{0.23\textwidth}|}
\hline
Differential equation & Form & True & 1\% noise & 10\% noise\\
\hline
Forced Duffing oscillator & $u_{tt}+\theta_1u_t+\theta_2u+\theta_3u^3=\gamma cos(\omega t)$ & $\theta=$[0.5, -1, 1, 0.42]  & \makecell{$\overline{\theta}=$[0.50,-0.99,1.00,0.42] \\ $cov=$[0.75,-0.33,0.29,0.83]\%}  & \makecell{$\overline{\theta}=$[0.49,-0.99,1.00,0.41] \\ $cov=$[3.16,-1.62,1.11,4.07]\%}  \\
\hline
Van der Pol oscillator & $u_{tt}+\theta_1u_t+\theta_2u^2u_t+\theta_3u=0$ & $\theta=$[-8, 8, 1] & \makecell{$\overline{\theta}=$[-7.77,7.91,0.97] \\ $cov=$[-1.6,1.56,1.49]\%} & \makecell{$\overline{\theta}=$[-7.23,7.40,0.92] \\ $cov=$[-4.18,5.26,3.00]\%}  \\
\hline
Burgers equation & $u_t+\theta_1uu_x+\theta_2u_{xx}=0$ & $\theta=$[1, -0.1] & \makecell{$\overline{\theta}=$[0.99,-0.1] \\ $cov=$[0.23,-0.56]\%} & \makecell{$\overline{\theta}=$[0.99,-0.1] \\ $cov=$[0.51,-1.65]\%}  \\
\hline
Korteweg-De Vries equation & $u_t+\theta_1uu_x+\theta_2u_{xxx}=0$ & $\theta=$[6, 1] & \makecell{$\overline{\theta}=$[5.94,0.97] \\ $cov=$[0.87,2.34]\%} & \makecell{$\overline{\theta}=$[5.95,0.97] \\ $cov=$[0.55,1.87]\%}  \\
\hline
Kuramoto Sivashinsky equation & $u_t+\theta_1uu_x+\theta_2u_{xx}+\theta_3u_{xxxx}=0$ & $\theta=$[1, 1, 1] & \makecell{$\overline{\theta}=$[0.91,0.92,0.91] \\ $cov=$[0.20,0.23,0.22]\%} & \makecell{$\overline{\theta}=$[0.87,0.88,0.89] \\ $cov=$[0.79,0.86,1.08]\%}  \\
\hline
2D Wave equation & $u_{tt} = \theta_1u_{xx}+\theta_2u_{yy}$ & $\theta=$[1, 1] & \makecell{$\overline{\theta}=$[0.99,1.00] \\ $cov=$[1.56,0.23]\%} & \makecell{$\overline{\theta}=$[0.94,0.99] \\ $cov=$[2.31,0.44]\%}  \\
\hline
Navier-Stokes equation & $w_{t}+\theta_1w_{xx}+\theta_2w_{yy} +\theta_3uw_{x}+\theta_4vw_{y}=0$& $\theta=$[-0.01, -0.01, 1, 1] & \makecell{$\overline{\theta}=$[-0.011, -0.009, 0.97, 0.98] \\ $cov=$[1.01,2.02,0.13,0.15]\%} & \makecell{$\overline{\theta}=$[-0.011,-0.009,0.94,0.94] \\ $cov=$[1.02,4.25,0.59,0.67]\%}  \\
\hline
\end{tabular}}
\label{tab:results}
\end{table}

\subsection{The Van der Pol oscillator}
\begin{figure}[h]
\centering
\includegraphics[width=0.8\textwidth]{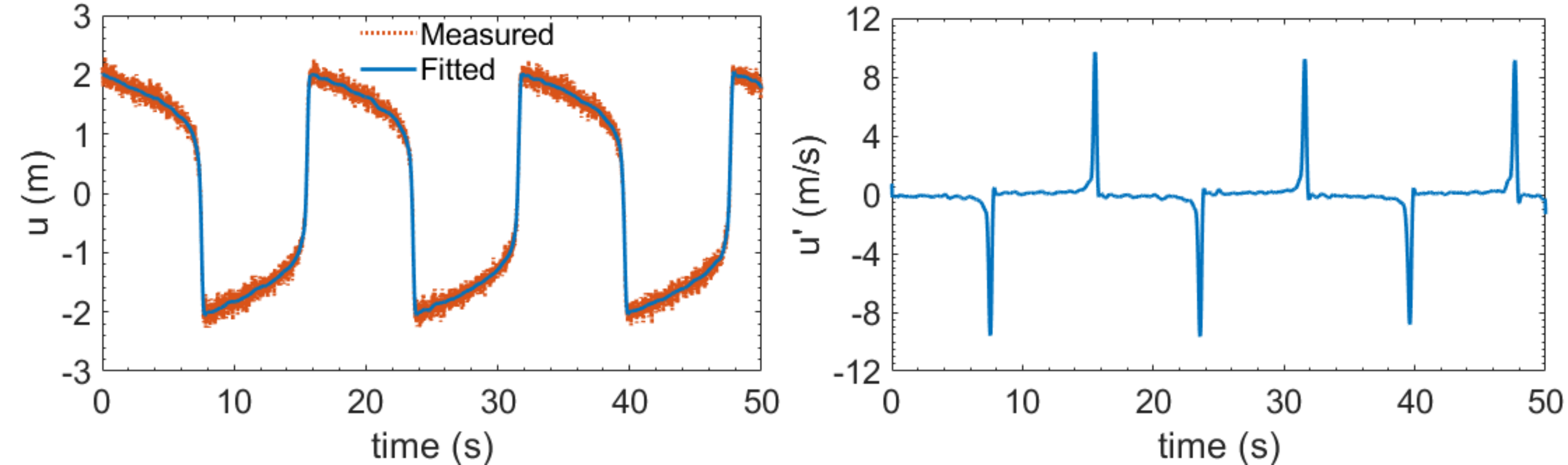}
\caption{Spline fitting to the measured data from Vanderpol oscillator and its first derivative derived from the proposed denoising algorithm}
\label{fig:vanderpol_denoise}
\end{figure}

\begin{figure}[h]
\centering
\includegraphics[width=0.8\textwidth]{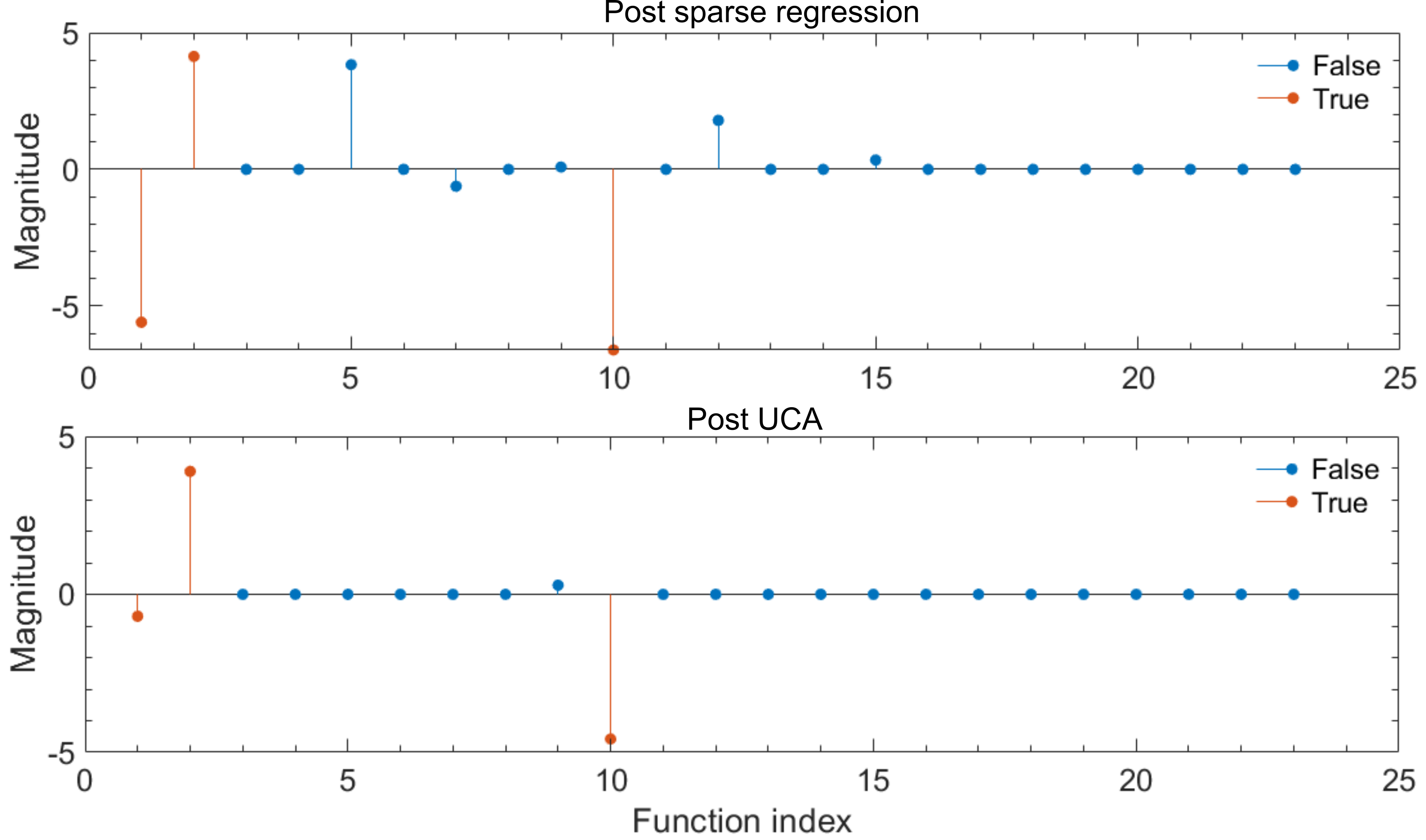}
\caption{Estimated parameter values for the Van der Pol oscillator post sparse regression and post UCA.}
\label{fig:vanderpol_intermediate}
\end{figure}

\begin{figure}[H]
\centering
\includegraphics[width=0.8\textwidth]{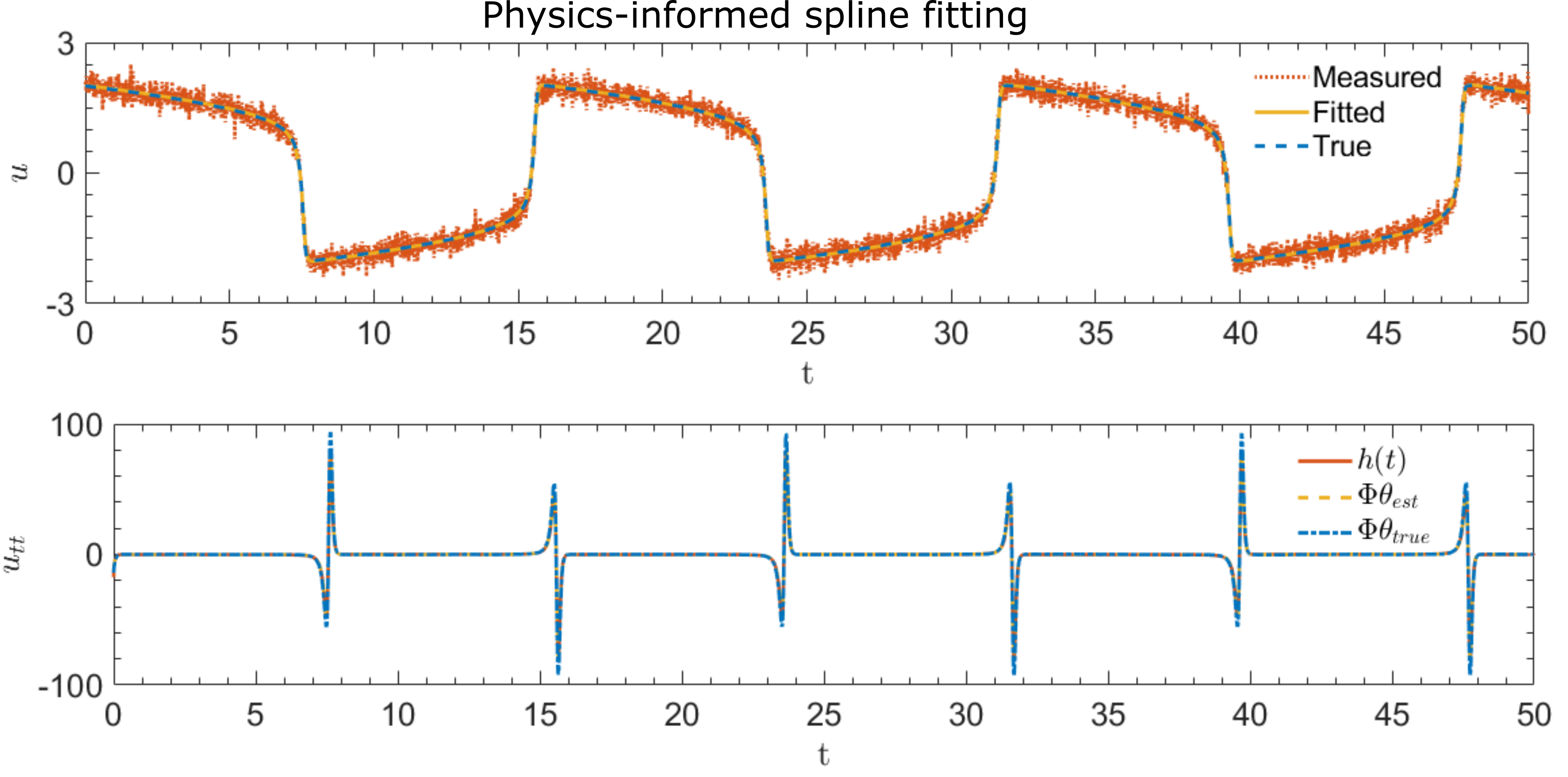}
\caption{The final spline fitting and comparison between the equation fitting obtained from true and identified parameters for the Van der Pol oscillator.}
\label{fig:vanderpol_identified}
\end{figure}

The Van der Pol oscillator represents a dynamic system with a stiff-differential equation of the following form:
\begin{equation}
    u_{tt}+\theta_1u_t+\theta_2u^2u_t+\theta_3u=0
\end{equation}
where $\theta = [\theta_1, \theta_2, \theta_3]$ are the parameters of the linear damping, nonlinear damping, and stiffness, respectively. The numerical solution for $\theta = [-8, 8, 1]$ was obtained for the oscillator's displacement response $u \in \mathbb{R}^{5000 \times 1}$, to which 10\% Gaussian noise was added to simulate the measured data. Figure \ref{fig:vanderpol_denoise} shows the signal fitting and first derivative after the denoising process. This oscillator shows a characteristic behavior of going through rapid changes at certain locations. This effect magnifies for the higher derivatives with high values at regions of rapid variations while remaining small otherwise.\\
The library comprises time-derivatives till 3\textsuperscript{rd} order, polynomial functions up to 5\textsuperscript{th} order, and cross-multiplication of derivatives and polynomial functions. As the derivatives differ by orders in magnitude due to the stiff behavior of the system, the factor $\bm{\Phi_{rms}}$ is set to 1 for the weighting vector $w$ in equation 9. The estimated coefficients post sparse regression and post UCA are shown in Figure \ref{fig:vanderpol_intermediate}. The function indexing corresponding to the Figure is provided in Table \ref{tab:function_index}. A few false functions (\#5$-u^3$, \#7$-u^5$, \#9$-uu_t$, \#12$-u^4u_t$, \#15$-u^2u_{tt}$) are present after sparse regression, with two of them (5, 12) having magnitudes comparable to the true functions. This happens because $u^3$ and $u^5$ share a high correlation with $u$, while $u^4u_t$ and $u^2u_{tt}$ share a high correlation with $u^2u_t$, which are the terms in the governing equation. These four functions mimic the behavior of the true functions and result in a high magnitude of its coefficients. Functions 5, 7, 12, and 15, sharing high correlation with the true functions, are eliminated at the UCA stage. While the only remaining false function 9 is easily eliminated at the PISF stage. Once the form of the governing equation is identified, a final run of the PISF is performed for the parameter estimation. For non-stiff differential equations, using higher-order splines does not affect the accuracy of parameter estimation. But for stiff differential equations like in this example, lower-order splines are recommended to capture the rapid transitions with better precision. Since the degree requirement for the splines was more for the proposed framework, parameter estimation can be performed using lower-order splines. Figure \ref{fig:vanderpol_identified} shows the spline fitting and the fitting of the governing equation for the Van der Pol oscillator. The fit quality is excellent, given the amount of noise in the measured data and the difficulty in capturing the stiff behavior of the oscillator.

\begin{table}
\centering
\caption{Functions in the library for Van der Pol oscillator}
\begin{tabular}{lcccc}
\hline
Functions & $u$ & $\frac{\partial^i u}{\partial t^i}$ & $u^i$ & $u^i\frac{\partial^j u}{\partial t^j}$ \\
Index & 1 & 2-3 & 4-7 & 9-24 \\
\hline
\end{tabular}
\label{tab:function_index}
\end{table}

\subsection{The forced duffing oscillator}
\begin{figure}[H]
\centering
\includegraphics[width=0.8\textwidth]{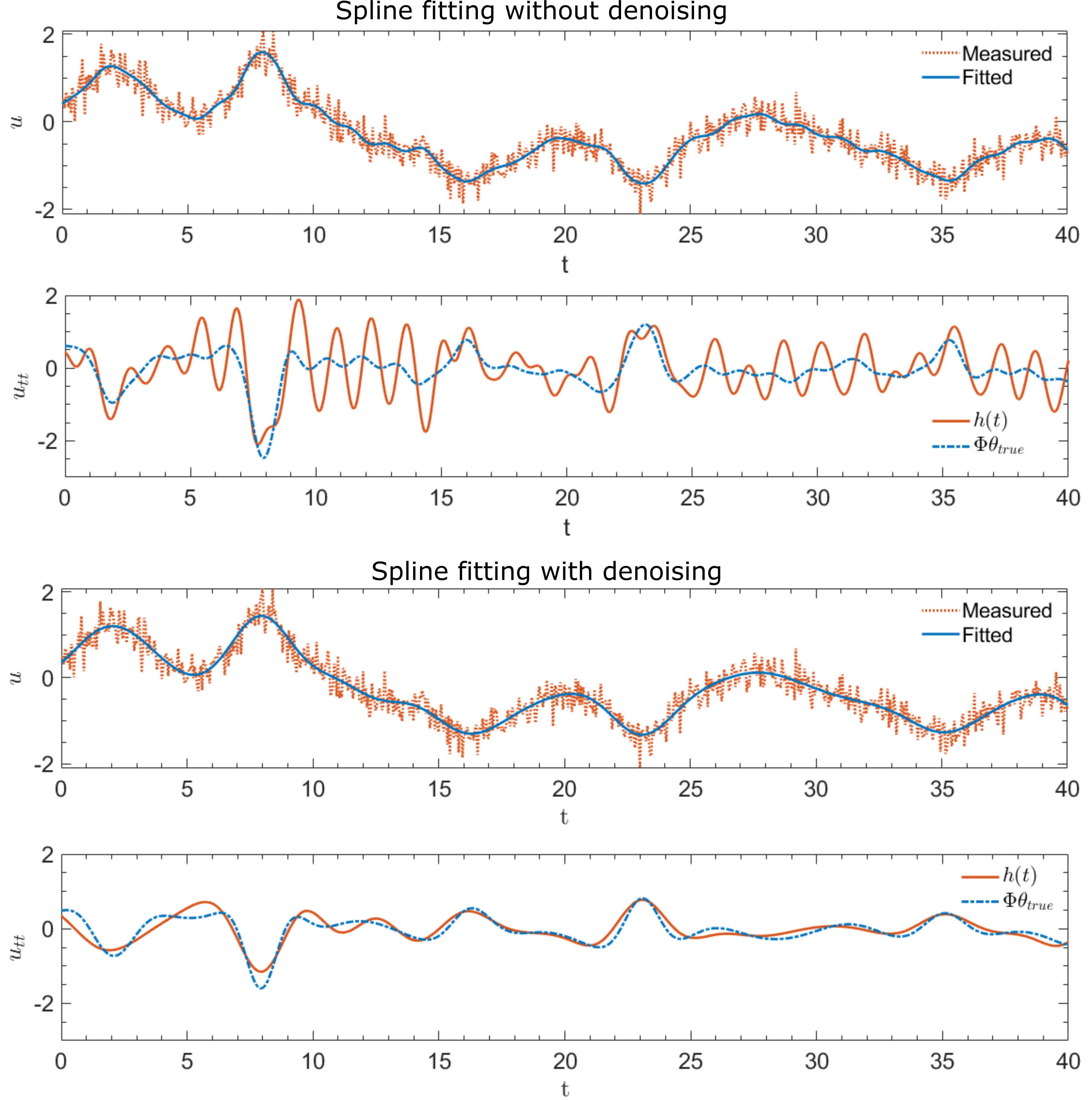}
\caption{Spline fitting with and without denoising, along with the fitting of the governing equation using true functions and parameter values.}
\label{fig:duffing_comparison}
\end{figure}

The governing equation of the Duffing oscillator with a cosine forcing function is given as follows:
\begin{equation}
    u_{tt}+\theta_1u_t+\theta_2u+\theta_3u^3=\gamma cos(\omega t)
\end{equation}
where $\theta = [\theta_1,\theta_2,\theta_3,\gamma]$ are the parameters of the linear damping, linear stiffness, nonlinear cubic stiffness, and the scaling factor of the forcing function, respectively. For $\theta = [0.5,-1,1,0.42]$, and $\omega=1$, the numerical solution of the ODE is obtained for $t=[0,200]$ seconds with a time step of 0.05 seconds, to which 10\% Gaussian noise (SNR=10) is added to simulate measured data. For equation discovery, $u_{tt}$ is chosen as $h(t)$ in equation 4. Figure \ref{fig:duffing_comparison} shows the fitting of B-splines to the data with/without smoothing (from 0-40 seconds) to illustrate the effect of noise by comparing $h_{t}$ and $\bm{\Phi\theta_{true}}$ in equation 4. The noise in the data causes significant fluctuations and mismatch between the $h(t)$ and $\bm{\Phi\theta_{true}}$. This mismatch allows false functions to contribute significantly during the identification process, making it hard to separate them from true functions. Denoising significantly improves this situation (see Figure \ref{fig:duffing_comparison}) and plays a key role throughout the identification process.

\begin{figure}[h]
\centering
\includegraphics[width=0.8\textwidth]{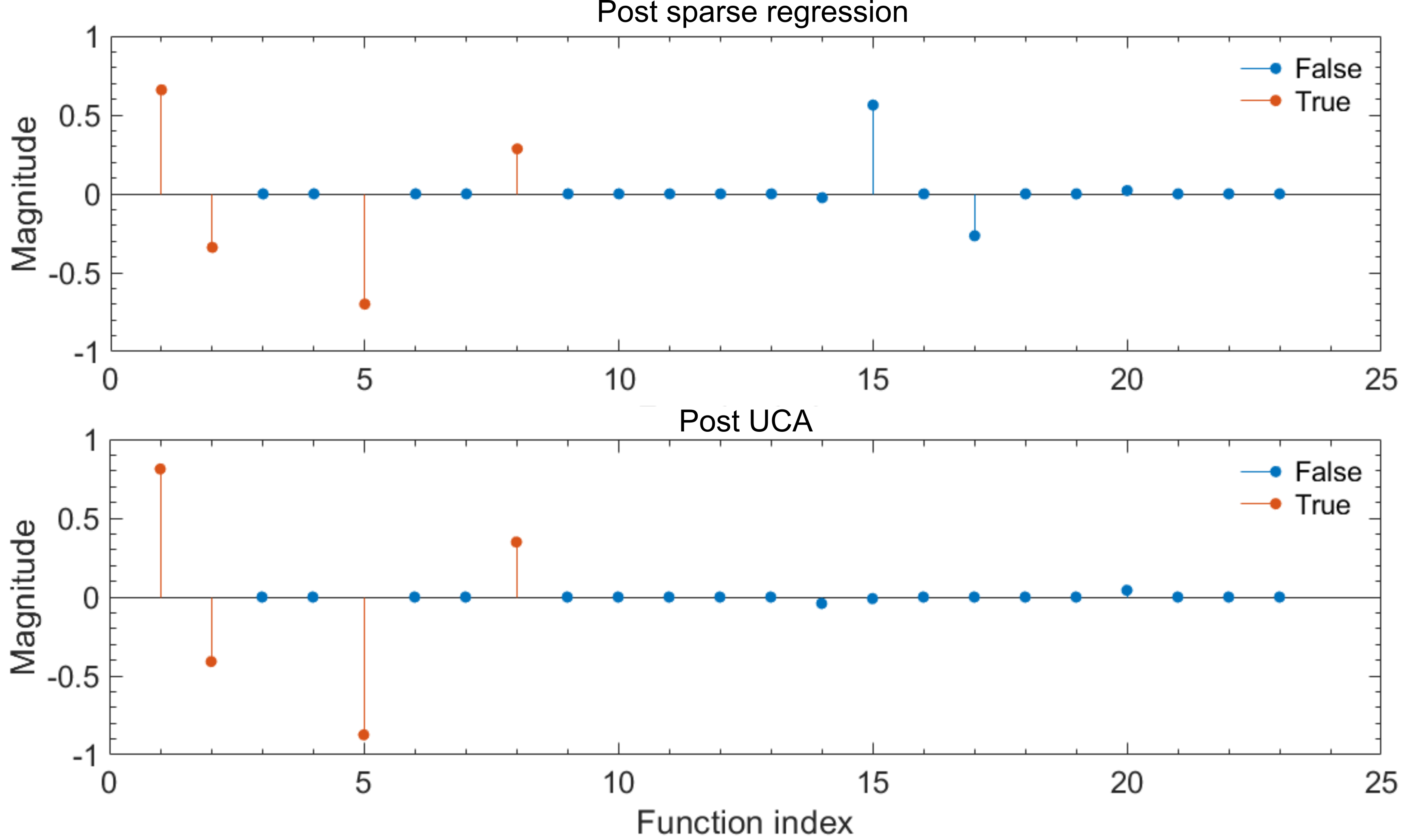}
\caption{Estimated parameter values for the Duffing oscillator post sparse regression and post UCA.}
\label{fig:duffing_intermediate}
\end{figure}

\begin{figure}[h]
\centering
\includegraphics[width=0.8\textwidth]{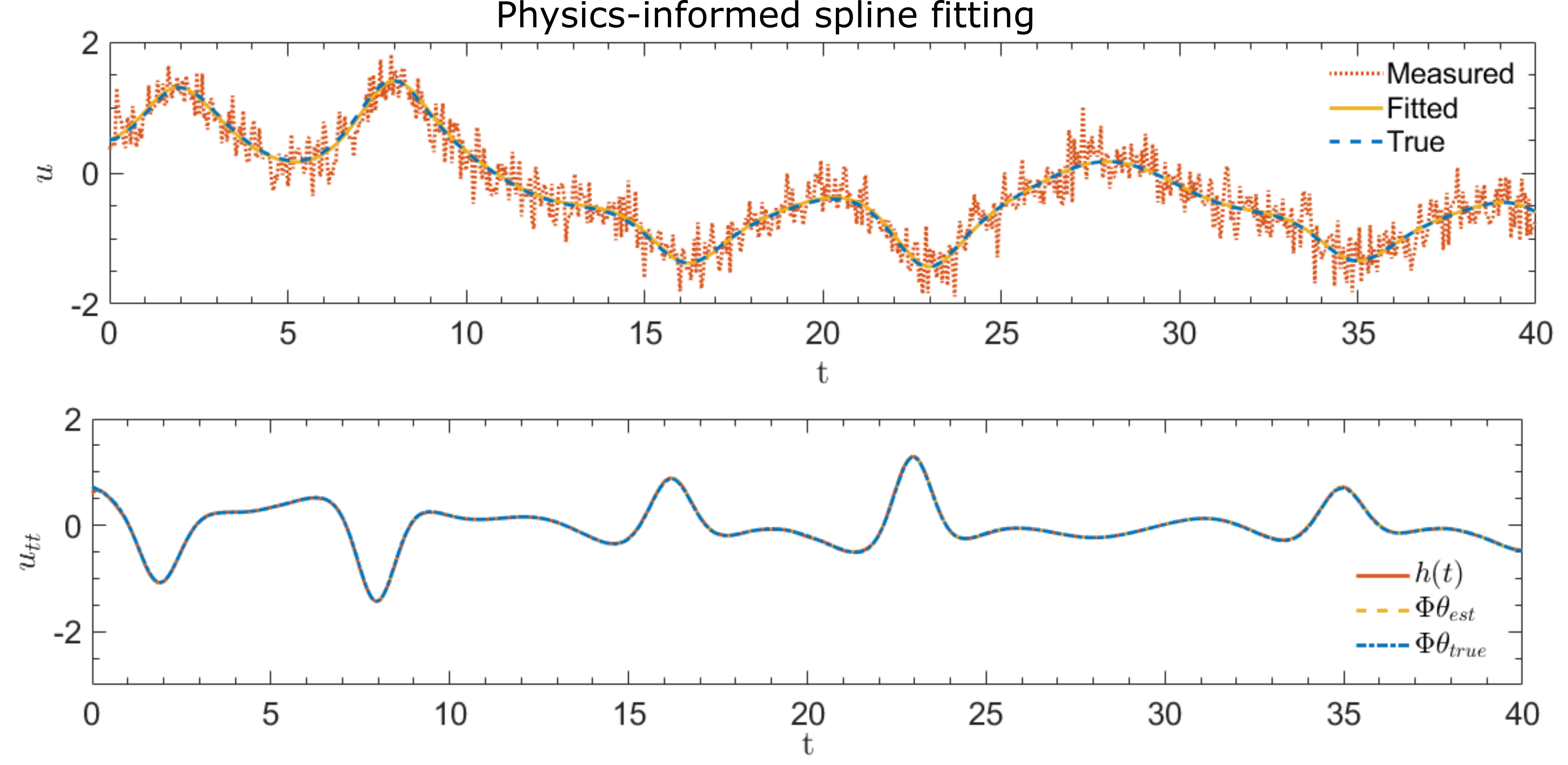}
\caption{The final spline fitting and comparison between the equation fitting obtained from true and identified parameters for the Duffing oscillator.}
\label{fig:duffing_identified}
\end{figure}

Since $u_{ttt}$ has been included in the library, $i_{max}=6$. The library $\bm{\Phi\theta_{true}}$ is formed of the same functions used for the Van der Pol oscillator. The estimated coefficients post sparse regression and post UCA are shown in Figure \ref{fig:duffing_intermediate}. During sparse regression, the coefficients of a couple of false functions (index 15 and 17) remain significantly high (even higher than two of the true functions), which maintains them as candidate functions for the governing equation. These functions get eliminated during the UCA stage, which was otherwise not possible to eliminate using criteria just based on low contribution. Also, functions 14 and 20 survived elimination during the sparse regression, which might be due to the threshold value being less than optimal. Choosing the correct threshold value has always been critical for accurate identification using sparse regression. In the proposed framework, if false functions with a small magnitude are retained due to sub-optimal thresholding, they get eliminated during the subsequent stages of UCA and PISF, which adds to the robustness of the proposed framework. \\
Figure \ref{fig:duffing_identified} shows the final data fitting and the comparison between $h_{t}$ and $\bm{\Phi\theta_{true}}$ for the true and identified parameters. The quality of fit is attributed to the physics-informed spline fitting that robustly fits equation 4 by adjusting the spline scaling factors ($\beta$). The identified form of the governing equation provides uniqueness, which guides the spline fitting and parameter estimation to high accuracy.

\subsection{Kuramoto-Sivashinsky equation}
\begin{figure}[h]
\centering
\includegraphics[width=0.8\textwidth]{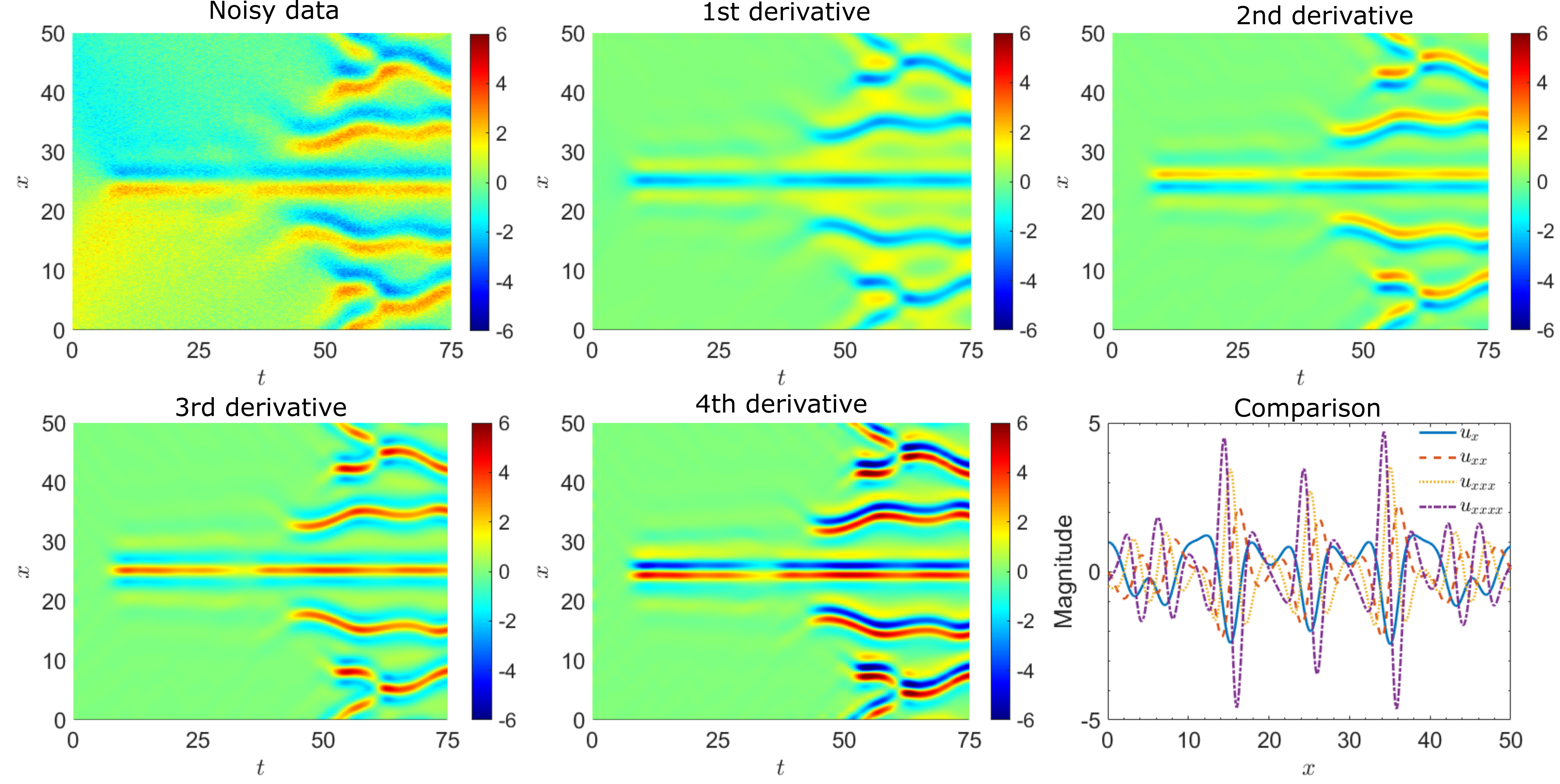}
\caption{Spatial derivatives of the measured noisy data (KS equation) up to the 4th order from the proposed denoising algorithm.}
\label{fig:kura_derivatives}
\end{figure}

The Kuramoto-Sivashinsky (KS) equation used in fluid dynamics applications is a fourth-order PDE of the following form:
\begin{equation}
    u_t+\theta_1uu_x+\theta_2u_{xx}+\theta_3u_{xxxx}=0
\end{equation}
where $\theta=[\theta_1, \theta_2, \theta_3]$ are the parameters of the equation with all of them equal to 1. The solution of the PDE is numerically obtained for $u \in \mathbb{R}^{512 \times 192}$ \cite{rudy2017data} to which 10\% Gaussian noise is added to simulate the measured data. This equation involves a fourth-order derivative which is especially difficult to calculate accurately even in the presence of small noise. Figure \ref{fig:kura_derivatives} shows the derivatives up to the fourth order calculated from the proposed method. It is impressive to see the smoothness of the derivatives irrespective of their order. Usually, the numerical differentiation of noisy signals for higher-order derivatives magnifies the effect of noise and the value of the calculated derivatives varies in a huge range as compared to the range of the measured signal. In Figure \ref{fig:kura_derivatives}, the range of the measured signal (with 10\% noise) is from -4 to 4 while its fourth derivative varies from -6 to 6, this gives evidence of the efficacy of the proposed strategy based on analytical differentiation and L\textsubscript{1}-regularization. The quality of the estimated derivatives makes the system identification possible for such a high-degree PDE.\\

\begin{figure}[h]
\centering
\includegraphics[width=0.8\textwidth]{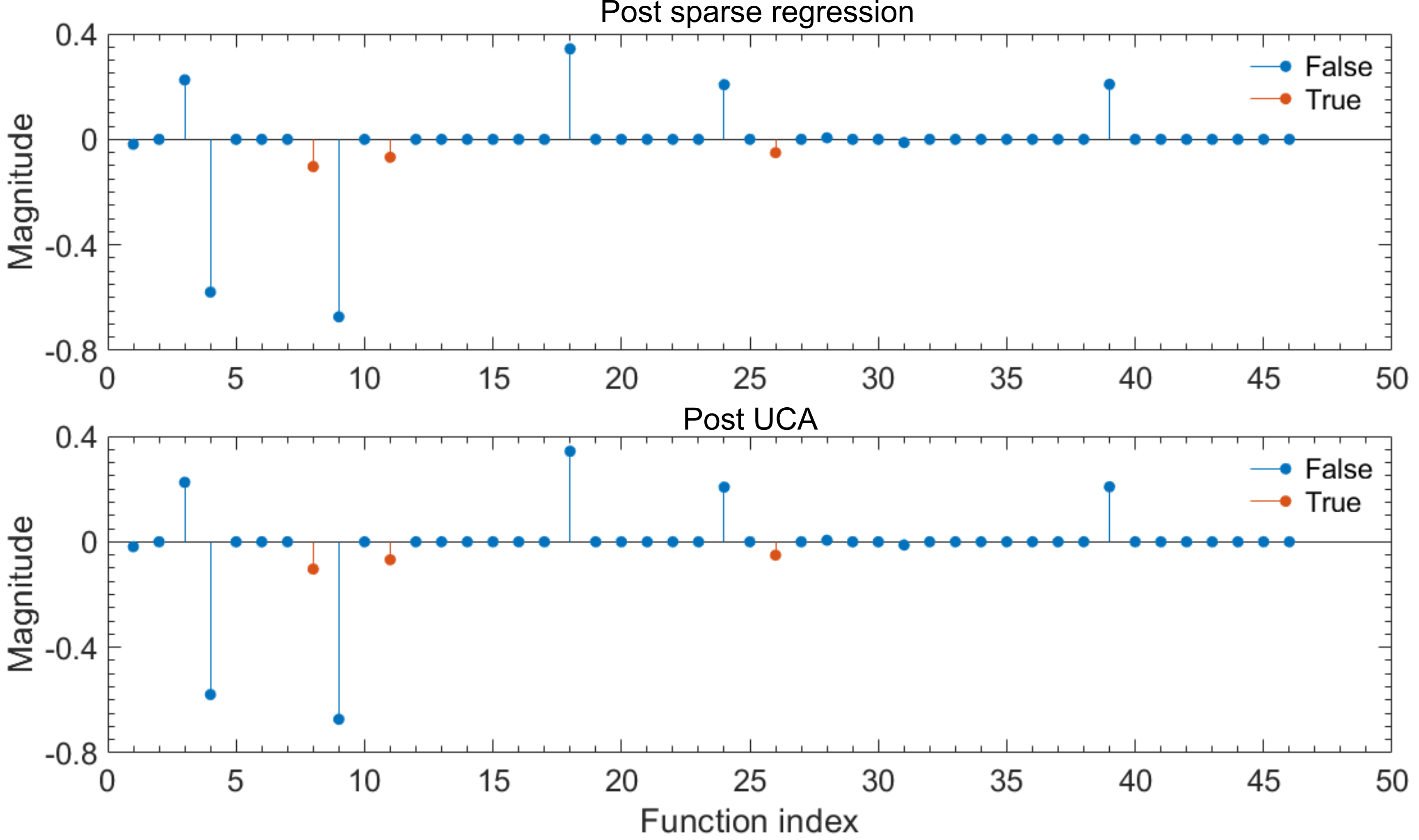}
\caption{Estimated parameter values for the KS equation post sparse regression and post UCA.}
\label{fig:kura_intermediate}
\end{figure}

\begin{figure}[h]
\centering
\includegraphics[width=0.6\textwidth]{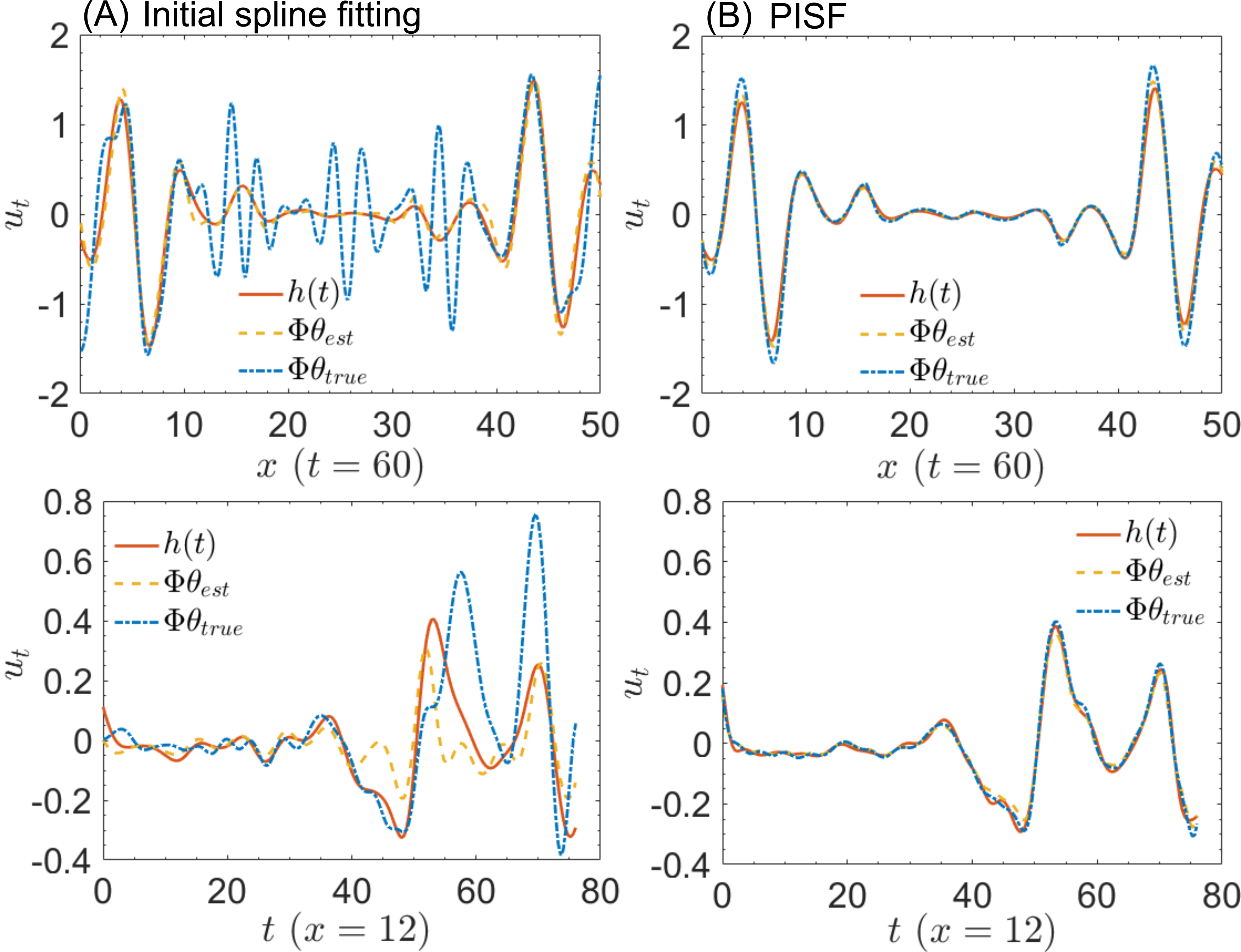}
\caption{Comparison between the governing equation fitting obtained from true and estimated parameters at $t=60$ and $x=12$ for a) Initial spline fitting, and b) PISF}
\label{fig:kura_compare}
\end{figure}

\begin{figure}[h]
\centering
\includegraphics[width=\textwidth]{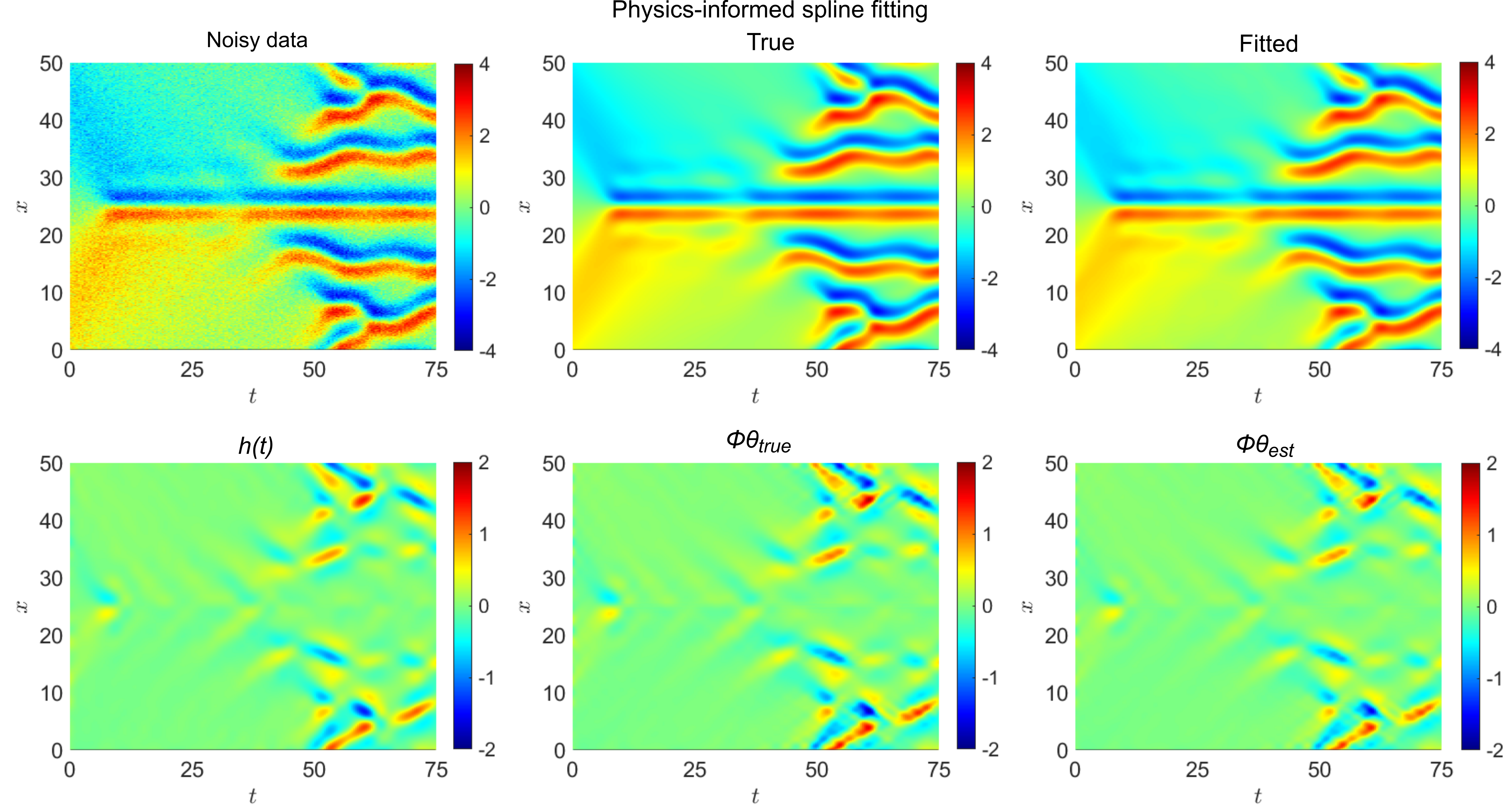}
\caption{The final spline fitting and comparison between the equation fitting obtained from true and identified parameters for the KS equation.}
\label{fig:kura_identified}
\end{figure}

The list of functions in $\bm{\Phi}$ includes time derivatives till the third order and space derivatives till the fourth order, mixed derivatives with a maximum order of three, polynomials up to the fifth degree, and cross-multiplication of polynomial and derivative functions. The estimated coefficients post sparse regression and post UCA are shown in Figure \ref{fig:kura_intermediate}. In this case, the coefficients of the true functions are smaller compared to the false functions that survived the sparse regression stage. Clearly, sparse regression as the sole strategy will not be able to identify the governing equation for this system. No function is eliminated at the UCA stage as none of the functions share a high cross-correlation. However, all the false functions are successfully eliminated at the PISF stage. Even though true functions had the lowest magnitude post the UCA stage, simultaneous spline fitting and satisfying the governing equation results in the increase of coefficient values of true functions and reduction for false functions. Figure \ref{fig:kura_compare} shows the comparison between the fitting of equation 4 obtained from initial denoising and PISF, at $x=12$ and $t=60$. The blue curve shows the estimated $h(t)$ for true values of $\bm{\beta}$. Ideally, the red and blue curves should coincide, but due to the presence of noise, the mismatch is significant for the initial spline fitting. On the other hand, the yellow curve that corresponds to the estimated $\bm{\beta}$ shows a better match. This is due to the abundance of functions in the library to choose from to satisfy equation 4. This explains why the magnitude of true functions is small and high for false functions post UCA stage in Figure \ref{fig:kura_intermediate}. Once the PISF stage is initiated, the spline fitting is modified in such a way that it simultaneously satisfies equation 4 and fits the measured data better. This guides the optimization towards the global minimum and sequentially eliminates the false functions. Once the form of the governing equation is identified the fitting of equation 4 is greatly improved, see Figure \ref{fig:kura_compare}. The identification of the correct form of a fourth-order equation with significant noise in the measured signal indicates the robustness of the proposed framework. Figure \ref{fig:kura_identified} shows the final spline fitting and compares the fitting of the governing equation.

\subsection{The Burgers equation}
\begin{figure}[h]
\centering
\includegraphics[width=0.7\textwidth]{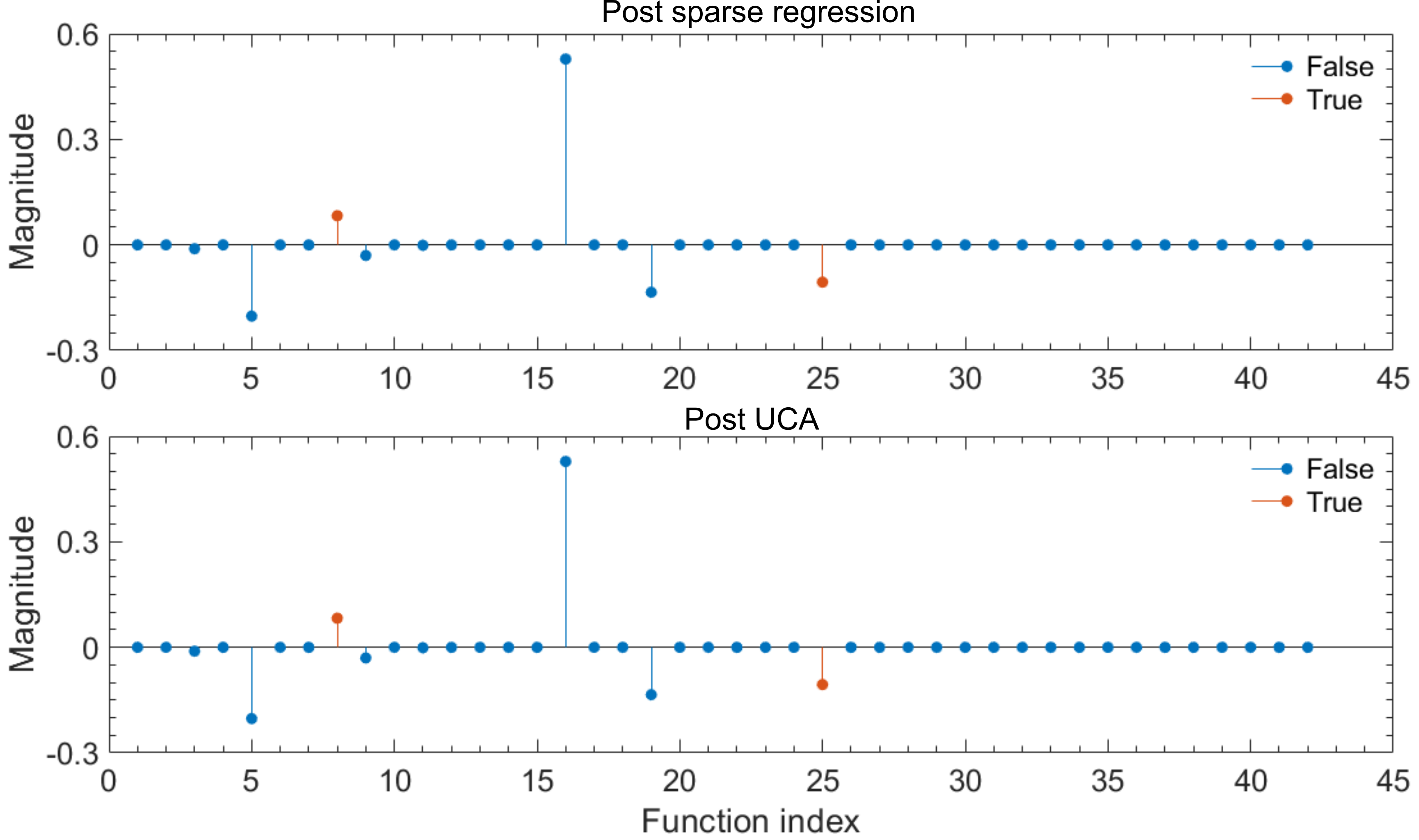}
\caption{Estimated parameter values for the Burger's equation post sparse regression and post UCA.}
\label{fig:burgers_intermediate}
\end{figure}

\begin{figure}[h]
\centering
\includegraphics[width=0.7\textwidth]{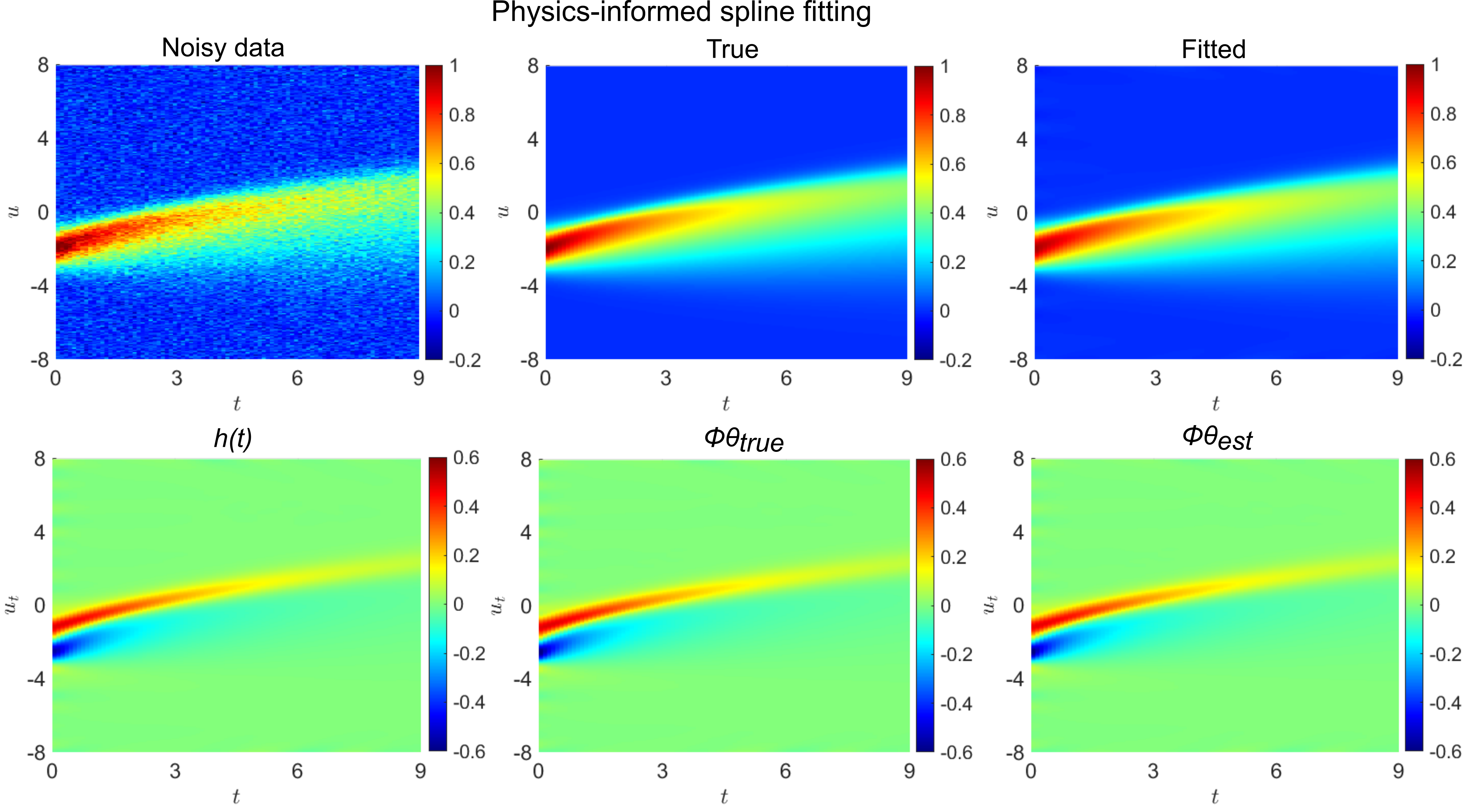}
\caption{The final spline fitting and comparison between the equation fitting obtained from true and identified parameters for Burger's equation.}
\label{fig:burgers_identified}
\end{figure}

The Burgers equation is a convection-diffusion equation with the PDE of the following form:
\begin{equation}
    u_t+\theta_1uu_x+\theta_2u_{xx}=0
\end{equation}
where $\theta=[\theta_1, \theta_2]$ are the parameters of equation of which $\theta_2$ depends on the viscosity of the fluid. For $\theta=[1, -0.1]$, the PDE is numerically solved to obtain the displacement response $u \in \mathbb{R}^{256 \times 96}$ \cite{rudy2017data} to which 10\% Gaussian noise is added to simulate the measured data. The list of functions in $\bm{\Phi}$ includes derivatives with respect to time and space till the third order, mixed derivatives with a maximum order of three, polynomials up to the fifth degree, and cross-multiplication of polynomial and derivative functions.\\
Figure \ref{fig:burgers_intermediate} shows the estimated coefficients post sparse regression and post UCA. Interestingly, three out of four false functions have coefficients greater than the true functions, indicating that sparse regression as the sole strategy will not work. Also, post UCA there isn't any further reduction because the functions do not share high correlation. All false functions are eliminated at the last stage of PISF which reflects the ability of simultaneous spline fitting and satisfying the governing equation to identify the correct form of the governing equation. Figure \ref{fig:burgers_identified} shows the final spline fitting and the fitting of equation 4 for the Burgers equation.

\subsection{Korteweg-De Vries equation}
\begin{figure}[h]
\centering
\includegraphics[width=0.7\textwidth]{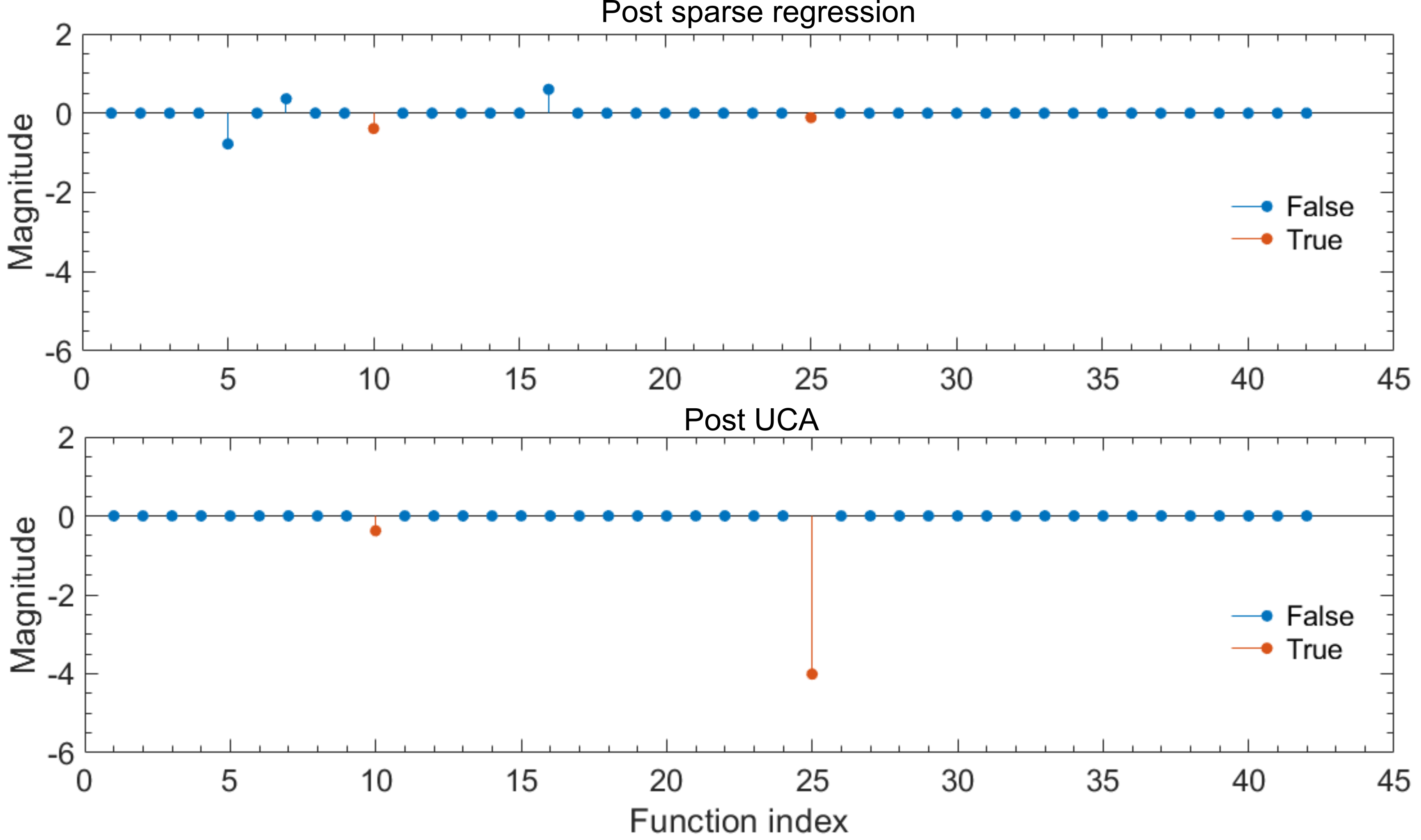}
\caption{Estimated parameter values for the KDV equation post sparse regression and post UCA.}
\label{fig:kdv_intermediate}
\end{figure}

\begin{figure}[h]
\centering
\includegraphics[width=0.8\textwidth]{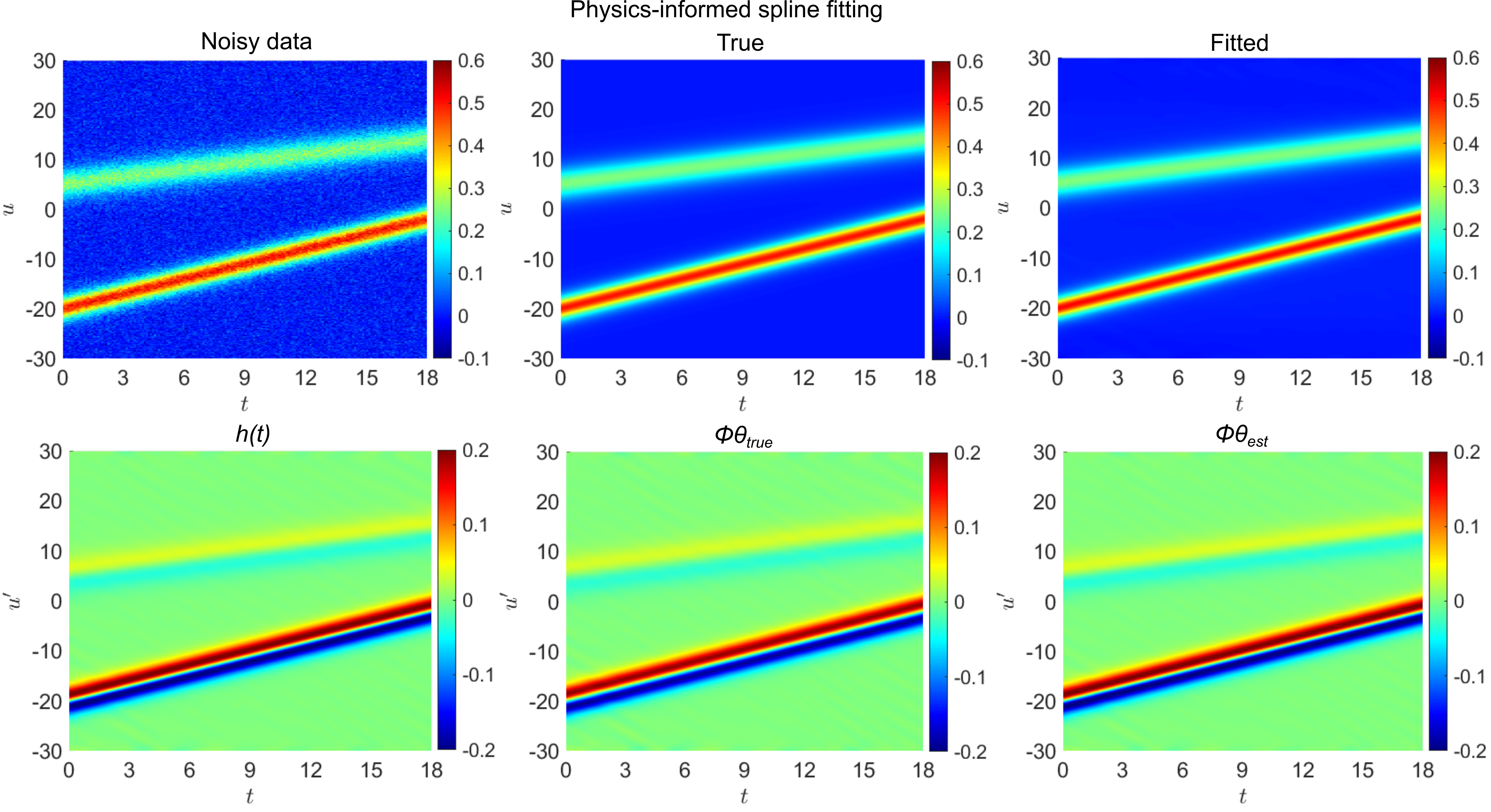}
\caption{The final spline fitting and comparison between the equation fitting obtained from true and identified parameters for the KDV equation.}
\label{fig:kdv_identified}
\end{figure}

The Korteweg-De Vries (KDV) equation is a nonlinear dispersive third-order PDE that models the wave on shallow water surfaces in space and time. The governing equation is of the following form:
\begin{equation}
    u_t+\theta_1uu_x+\theta_2u_{xxx}=0
\end{equation}
where $\theta=[\theta_1, \theta_2]$ are the parameters of the advection and dispersion terms, respectively. For $\theta=[6,1]$, the PDE is solved numerically to obtain the displacement response $u \in \mathbb{R}^{512 \times 192}$ \cite{rudy2017data} to which 10\% Gaussian noise is added to simulate the measured data. The dictionary used for this system is the same as the previously studied Burger's equation. Figure \ref{fig:kdv_intermediate} shows the estimated coefficients post sparse regression and post UCA. The effect of the presence of false functions is visible in this example case. The true functions which have a magnitude of 6 and 1, do not even exceed 0.3 in the presence of false functions. The primary reason for this behavior is the high correlation between all the functions that survived sparse regression. The minimum cross-correlation is above 0.85 for any combination of two functions. This causes the coefficients of false functions to be comparable/higher than true functions and significantly difficult to detect and eliminate. The proposed UCA algorithm successfully identifies the false functions and eliminates them to get the correct form of the governing equation. Since the minimum cross-correlation was high in this example, the value of the threshold had to be increased to identify the complementary functions. The final spline fitting and comparison of the governing equation is shown in Figure \ref{fig:kdv_identified}.

\subsection{Navier-Stokes equation}
\begin{figure}[h]
\centering
\includegraphics[width=0.8\textwidth]{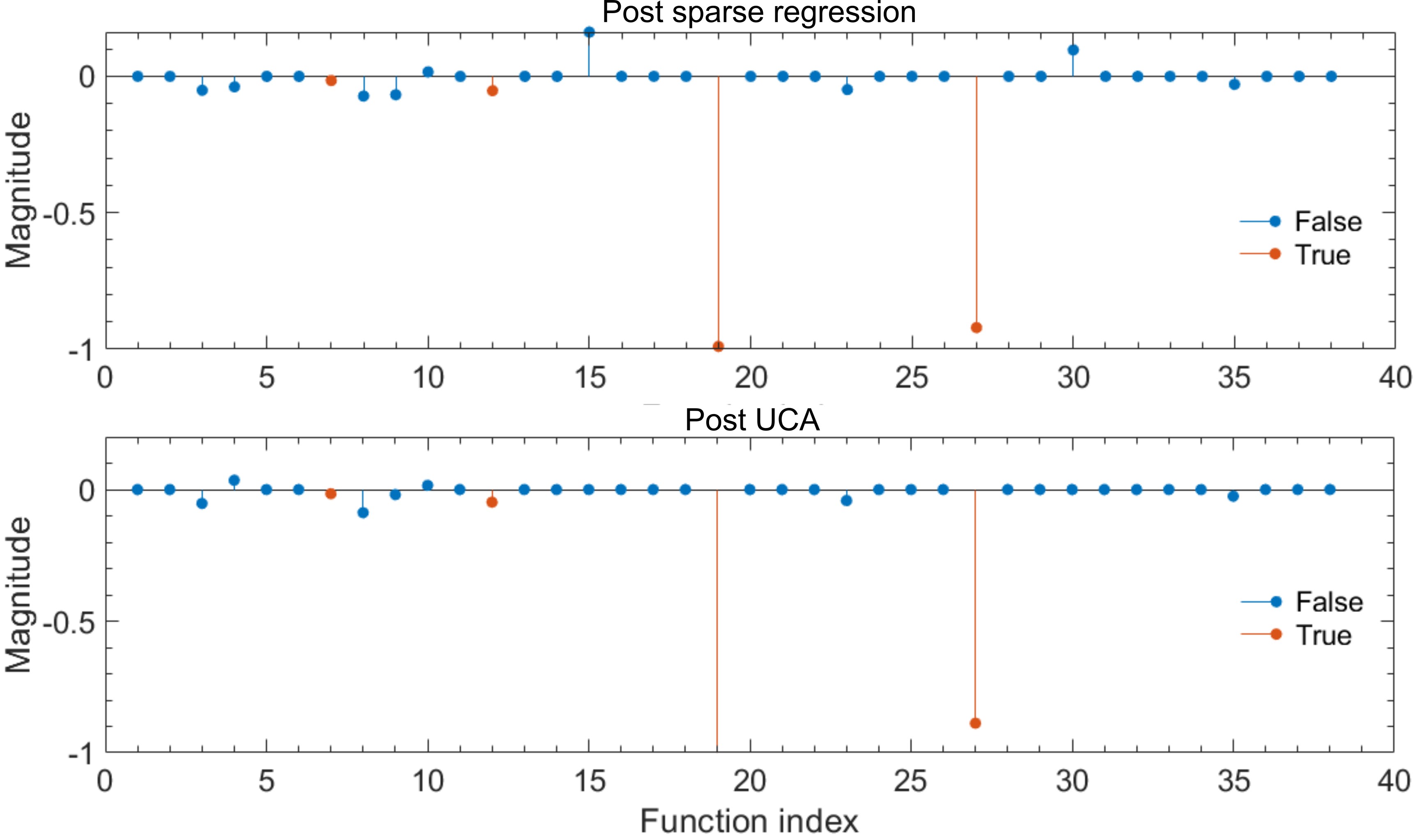}
\caption{Estimated parameter values for the Navier-Stokes equation post sparse regression and post UCA.}
\label{fig:navier_intermediate}
\end{figure}

\begin{figure}[h]
\centering
\includegraphics[width=\textwidth]{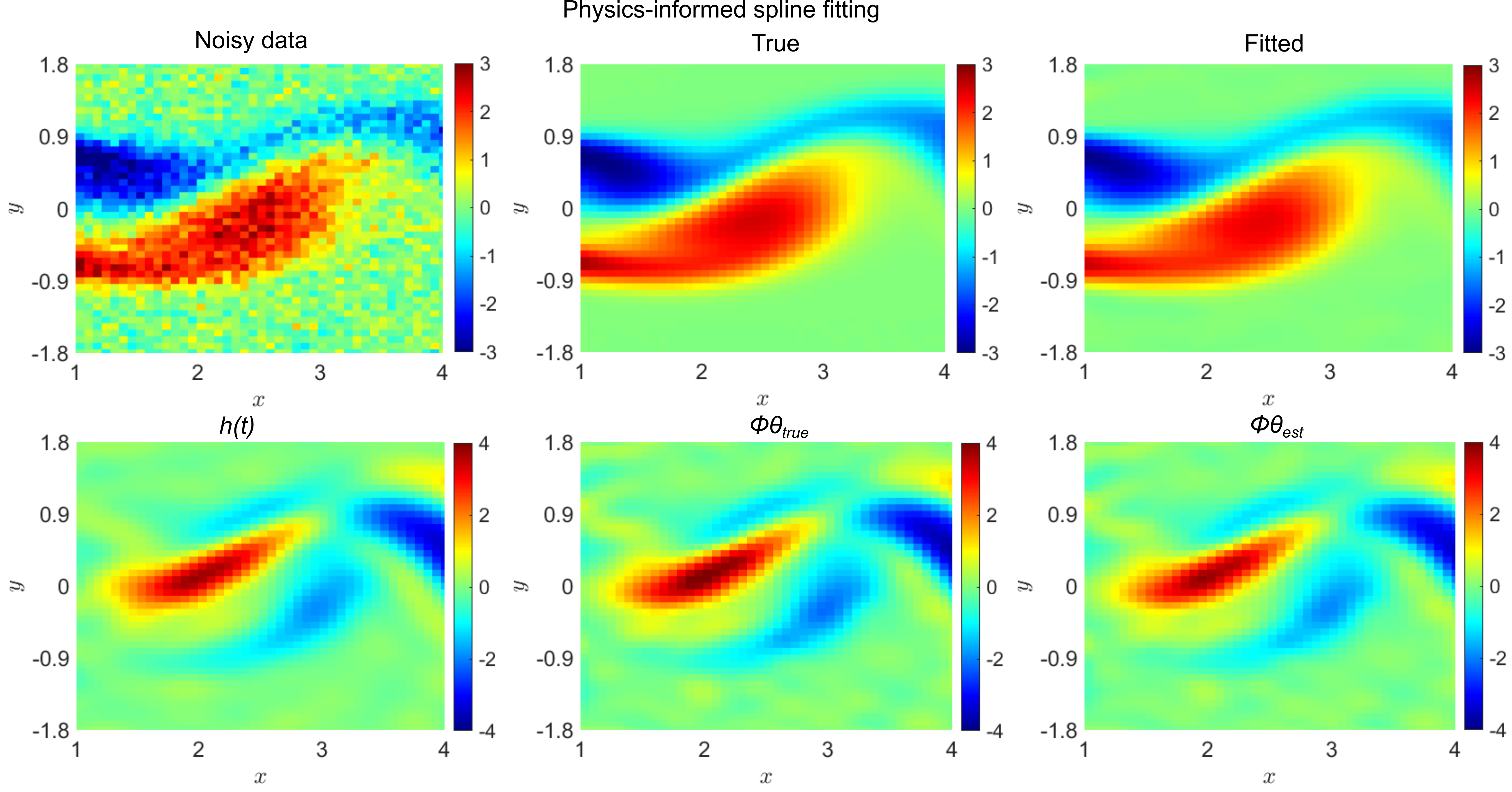}
\caption{The final spline fitting and comparison between the equation fitting obtained from true and identified parameters for the Navier-Stokes equation.}
\label{fig:navier_identified}
\end{figure}

The Navier-Stokes equation is a partial differential equation describing viscous fluid substances' motion. It has been applied to various phenomena in science and engineering, such as modeling the weather, ocean currents, water flow in pipes, airflow around a wing, blood flow, etc. The second-order PDE is described as follows:
\begin{equation}
    w_{t}+\theta_1w_{xx}+\theta_2w_{yy}+\theta_3uw_x+\theta_4vw_y=0
\end{equation}
where $u(x,y,t), v(x,y,t) \in \mathbb{R}^{48 \times 48 \times 64}$ are the velocity field data, and $w(x,y,t) \in \mathbb{R}^{48 \times 48 \times 64}$ is the vorticity data calculated from the velocity fields, and $\theta = [\theta_1, \theta_2, \theta_3, \theta_4]$ is the set of parameters. The PDE is numerically solved for $\theta = [0.01, 0.01, 1, 1]$ to get the velocity field and vorticity data \cite{raissi2019physics} to which 10\% noise is added to simulate the measured data. The dictionary is formed of derivatives in space and time till the second order, mixed derivatives, velocity fields, vorticity, and cross-multiplication of derivatives with velocities and vorticity. The estimated coefficients after the sparse-regression and post UCA are shown in Figure \ref{fig:navier_intermediate}. Two functions (19, 27) of the four true functions (7, 12, 19, 27) show significantly high magnitude, while the other two have magnitude smaller than even false functions. If thresholding is chosen as the sole strategy to get the sparse solution, then functions 15 and 30 will remain, while the true functions 7 and 12 will be eliminated, leading to an incorrect solution. In Figure \ref{fig:navier_intermediate} post UCA, the functions 15 and 30 that had the third and fourth highest magnitude are rightly eliminated, highlighting the effectiveness of the UCA algorithm. A few false functions remain at this stage, even with a higher magnitude than the true functions. These false functions are sequentially eliminated at the last stage of PISF, converging to the correct PDE with four terms. The final spline fitting and the governing equation for the Navier-Stokes equation at $t=3$ are shown in Figure \ref{fig:navier_identified}. In a few cases, the term $w_{xx}$ is replaced by $w_x$ during the identification process. The contribution of $w_{xx}$ is so small that the uniqueness of the term is absent, and the term $w_x$ captures the contribution from $w_{xx}$ equally well. For the cases when the contribution of $w_{xx}$ increases, it is expected that $w_{xx}$ should be identified each time.

\subsection{2D Wave equation}
\begin{figure}[h]
\centering
\includegraphics[width=0.7\textwidth]{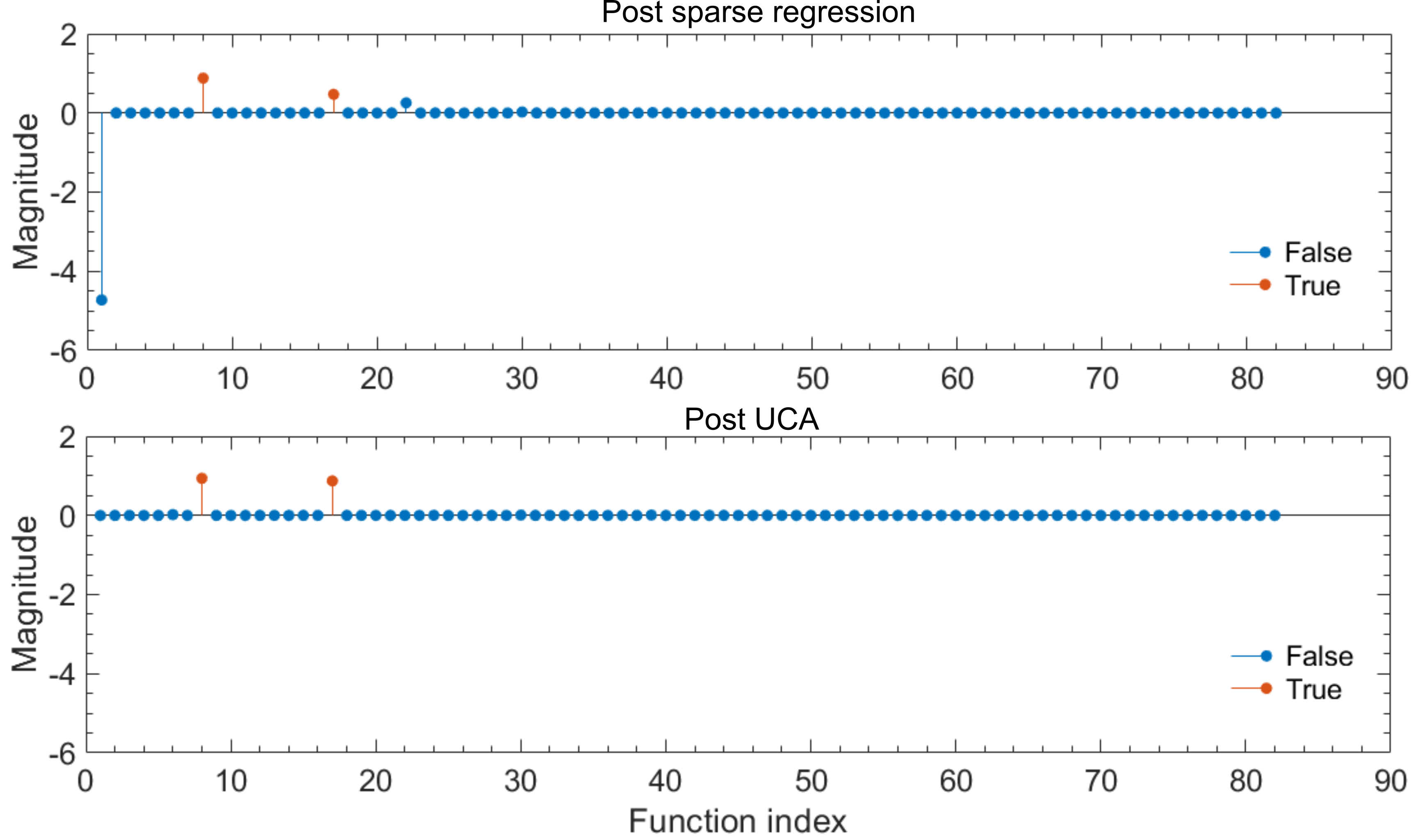}
\caption{Estimated parameter values for the 2D-Wave equation post sparse regression and post UCA.}
\label{fig:2dwave_intermediate}
\end{figure}

\begin{figure}[h]
\centering
\includegraphics[width=0.8\textwidth]{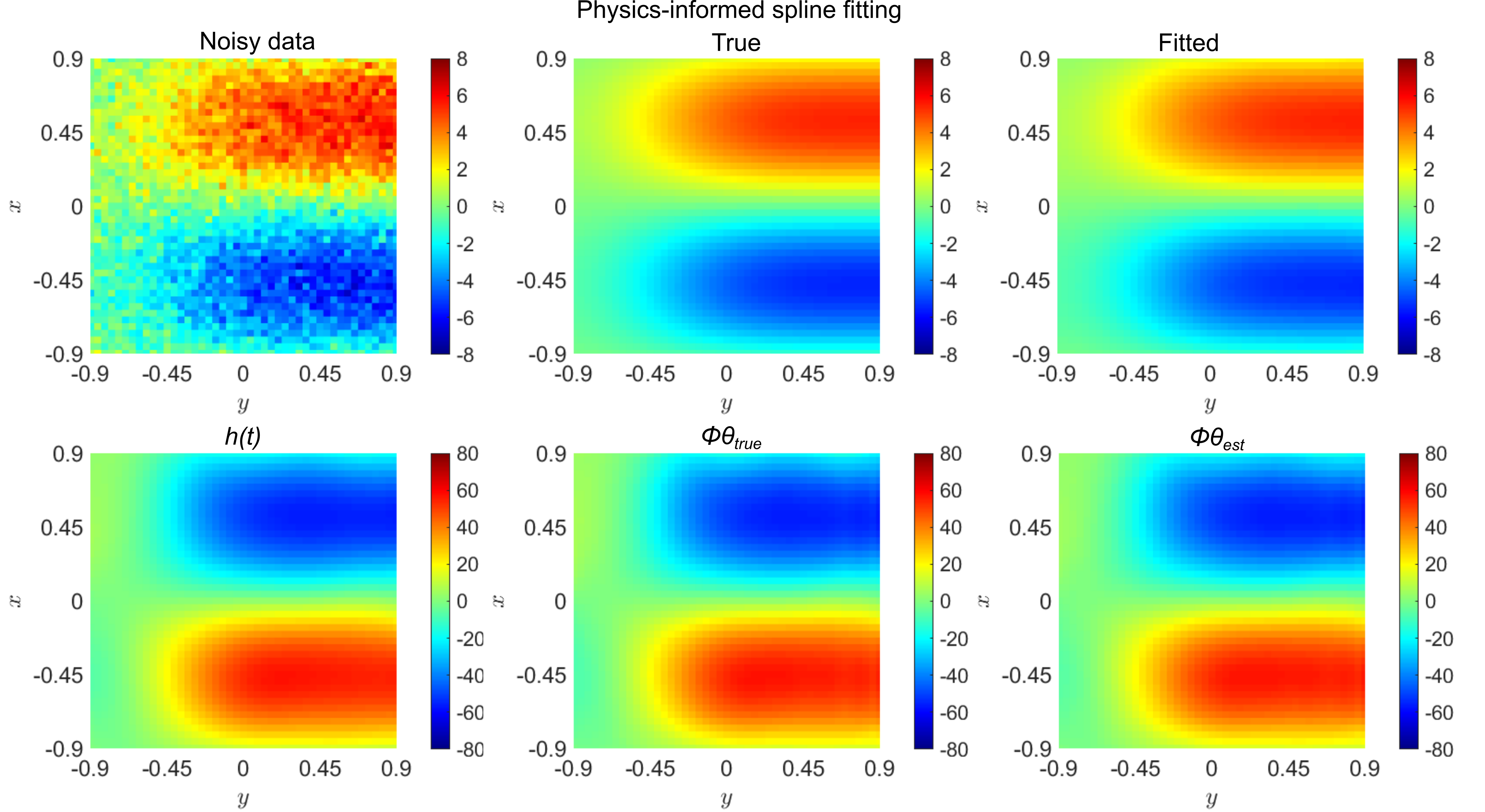}
\caption{The final spline fitting and comparison between the equation fitting obtained from true and identified parameters for the 2D-Wave equation.}
\label{fig:2dwave_identified}
\end{figure}

The wave equation is a second-order PDE describing waves according to the following equation:
\begin{equation}
    u_{tt} = \theta_1u_{xx}+\theta_2u_{yy}
\end{equation}
where $\theta = [\theta_1, \theta_2]$ are the parameters corresponding to the wave speed. For the example case, PDE is numerically solved for $\theta = [1, 1]$ to get the wave displacement response $u \in \mathbb{R}^{48 \times 48 \times 48}$ to which 10\% noise is added to simulate the measured data. The dictionary is formed in the same fashion as the previous examples, containing derivatives in each variable to the third degree, mixed derivatives, polynomials, and cross-multiplication of polynomial and derivative functions. The denoising is performed by simultaneously regularizing derivatives for the three variables. The estimated coefficients post sparse regression and post UCA are shown in Figure \ref{fig:2dwave_intermediate}. A few false functions also escaped the sparse regression in this example, which are effectively eliminated using the UCA algorithm. The rest of the false functions are removed in the PISF stage. The final spline fitting and equation 4 are shown in Figure \ref{fig:2dwave_identified}. 

\section{Conclusion}
The sparse system identification methods effectively discover the governing equation from noisy measured data. The existing methods struggle to identify the correct equation in the case of stiff and high-order differential equations for noisy data and highly correlated functions in the dictionary. In this study, we demonstrate a novel algorithm comprising three essential steps of sparse regression, UCA, and PISF for sparse system identification. The proposed method was tested on various ODE/PDE consisting of three-dimensional PDEs, fourth-order PDE, and stiff/non-stiff systems. For each case, the method discovers the correct equation even in the presence of high noise in the measured data. The parameter estimation for each ODE/PDE is also accurate and has a small coefficient of variation, indicating the method's robustness to the noise.\\
The presence of noise is one of the major factors that challenge the identification of the correct equation. A novel SRDD algorithm is demonstrated based on the sequential regularization of the derivatives of the signals that effectively removes noise from the data without losing information on the system response, as seen in the case of the Van der Pol oscillator. Further, using B-spline functions enables analytical differentiation of the fitted signal to obtain accurate derivatives to the desired order, as seen in the case of the Kuramoto-Sivashinsky equation.\\
Along with noise, choosing functions that share a high correlation with true functions poses a challenge in discovering the correct equation. The introduction of UCA successfully eliminates such functions to reduce the number of candidates in the library. In the case of Van der Pol oscillator and KDV equation, it is observed that functions with significantly high magnitude are also eliminated with the proposed UCA algorithm. In more challenging problems such as the KS equation, the true functions have the smallest coefficients post sparse regression and UCA. The PISF updates the spline fitting in such a way that the true coefficients increase in magnitude and false functions get eliminated sequentially, converging to the true PDE.\\
The proposed method is a useful tool for finding the guiding dynamic equation underlying the measured data. The denoising algorithm, followed by the proposed three-step algorithm, takes advantage of the physics-informed fitting of the measured data to converge to the correct PDE. The simultaneous data fitting and satisfying the governing equation gradually leads the optimization to the global minimum which is otherwise extremely hard to reach for certain systems. The applicability of the method has been demonstrated for systems across fields of civil, mechanical, and fluid dynamics but is not limited to them. The identification method is suitable for any system with the available field measurement data and is expected to correctly discover the hidden underlying dynamic equation successfully. Future advances can be made toward the model discovery of coupled ODEs and PDEs. This will be beneficial in understanding systems where parts of it behave differently, but the interaction between them impacts their behavior and the overall system.

\section*{Acknowledgments}
The authors are grateful to the Science and Engineering Research Board of India for the financial support under fund number R96704.

%Bibliography
\bibliographystyle{unsrt}  
\bibliography{references}

\begin{thebibliography}{10}

\bibitem{lai2019semi}
Zhilu Lai and Satish Nagarajaiah.
\newblock Semi-supervised structural linear/nonlinear damage detection and characterization using sparse identification.
\newblock {\em Structural Control and Health Monitoring}, 26(3):e2306, 2019.

\bibitem{bhowmick2023physics}
Sutanu Bhowmick and Satish Nagarajaiah.
\newblock Physics-guided identification of euler--bernoulli beam pde model from full-field displacement response with simultaneous basis function approximation and parameter estimation (snape).
\newblock {\em Engineering Structures}, 289:116231, 2023.

\bibitem{jimenez2020maintenance}
Alfredo~Arcos Jim{\'e}nez, Long Zhang, Carlos Quiterio~G{\'o}mez Mu{\~n}oz, and Fausto Pedro~Garc{\'\i}a M{\'a}rquez.
\newblock Maintenance management based on machine learning and nonlinear features in wind turbines.
\newblock {\em Renewable Energy}, 146:316--328, 2020.

\bibitem{alber2019integrating}
Mark Alber, Adrian Buganza~Tepole, William~R Cannon, Suvranu De, Salvador Dura-Bernal, Krishna Garikipati, George Karniadakis, William~W Lytton, Paris Perdikaris, Linda Petzold, et~al.
\newblock Integrating machine learning and multiscale modeling—perspectives, challenges, and opportunities in the biological, biomedical, and behavioral sciences.
\newblock {\em NPJ digital medicine}, 2(1):115, 2019.

\bibitem{sharma2022hybrid}
Niket Sharma and YA~Liu.
\newblock A hybrid science-guided machine learning approach for modeling chemical processes: A review.
\newblock {\em AIChE Journal}, 68(5):e17609, 2022.

\bibitem{wang2021neural}
Xueli Wang, Derui Ding, Hongli Dong, and Xian-Ming Zhang.
\newblock Neural-network-based control for discrete-time nonlinear systems with input saturation under stochastic communication protocol.
\newblock {\em IEEE/CAA Journal of Automatica Sinica}, 8(4):766--778, 2021.

\bibitem{bruder2019nonlinear}
Daniel Bruder, C~David Remy, and Ram Vasudevan.
\newblock Nonlinear system identification of soft robot dynamics using koopman operator theory.
\newblock In {\em 2019 International Conference on Robotics and Automation (ICRA)}, pages 6244--6250. IEEE, 2019.

\bibitem{klus2020data}
Stefan Klus, Feliks N{\"u}ske, Sebastian Peitz, Jan-Hendrik Niemann, Cecilia Clementi, and Christof Sch{\"u}tte.
\newblock Data-driven approximation of the koopman generator: Model reduction, system identification, and control.
\newblock {\em Physica D: Nonlinear Phenomena}, 406:132416, 2020.

\bibitem{marchesiello2008time}
Stefano Marchesiello and Luigi Garibaldi.
\newblock A time domain approach for identifying nonlinear vibrating structures by subspace methods.
\newblock {\em Mechanical Systems and Signal Processing}, 22(1):81--101, 2008.

\bibitem{ayala2020nonlinear}
Helon Vicente~Hultmann Ayala, Didace Habineza, Micky Rakotondrabe, and Leandro dos Santos~Coelho.
\newblock Nonlinear black-box system identification through coevolutionary algorithms and radial basis function artificial neural networks.
\newblock {\em Applied Soft Computing}, 87:105990, 2020.

\bibitem{yu2019system}
Dongmin Yu, Yong Wang, Huanan Liu, Kittisak Jermsittiparsert, and Navid Razmjooy.
\newblock System identification of pem fuel cells using an improved elman neural network and a new hybrid optimization algorithm.
\newblock {\em Energy Reports}, 5:1365--1374, 2019.

\bibitem{li2020fourier}
Zongyi Li, Nikola Kovachki, Kamyar Azizzadenesheli, Burigede Liu, Kaushik Bhattacharya, Andrew Stuart, and Anima Anandkumar.
\newblock Fourier neural operator for parametric partial differential equations.
\newblock {\em arXiv preprint arXiv:2010.08895}, 2020.

\bibitem{silvestrini2022deep}
Stefano Silvestrini and Mich{\`e}le Lavagna.
\newblock Deep learning and artificial neural networks for spacecraft dynamics, navigation and control.
\newblock {\em Drones}, 6(10):270, 2022.

\bibitem{wu2019deep}
Rih-Teng Wu and Mohammad~R Jahanshahi.
\newblock Deep convolutional neural network for structural dynamic response estimation and system identification.
\newblock {\em Journal of Engineering Mechanics}, 145(1):04018125, 2019.

\bibitem{de2016randomized}
Erick De~la Rosa and Wen Yu.
\newblock Randomized algorithms for nonlinear system identification with deep learning modification.
\newblock {\em Information Sciences}, 364:197--212, 2016.

\bibitem{raissi2018hidden}
Maziar Raissi and George~Em Karniadakis.
\newblock Hidden physics models: Machine learning of nonlinear partial differential equations.
\newblock {\em Journal of Computational Physics}, 357:125--141, 2018.

\bibitem{bhowmick2023data}
Sutanu Bhowmick, Satish Nagarajaiah, and Anastasios Kyrillidis.
\newblock Data-and theory-guided learning of partial differential equations using simultaneous basis function approximation and parameter estimation (snape).
\newblock {\em Mechanical Systems and Signal Processing}, 189:110059, 2023.

\bibitem{raissi2019physics}
Maziar Raissi, Paris Perdikaris, and George~E Karniadakis.
\newblock Physics-informed neural networks: A deep learning framework for solving forward and inverse problems involving nonlinear partial differential equations.
\newblock {\em Journal of Computational physics}, 378:686--707, 2019.

\bibitem{lai2021structural}
Zhilu Lai, Charilaos Mylonas, Satish Nagarajaiah, and Eleni Chatzi.
\newblock Structural identification with physics-informed neural ordinary differential equations.
\newblock {\em Journal of Sound and Vibration}, 508:116196, 2021.

\bibitem{pal2024sparsity}
Ashish Pal and Satish Nagarajaiah.
\newblock Sparsity promoting algorithm for identification of nonlinear dynamic system based on unscented kalman filter using novel selective thresholding and penalty-based model selection.
\newblock {\em Mechanical Systems and Signal Processing}, 212:111301, 2024.

\bibitem{zhang2017structural}
Chun Zhang, Jie-Zhong Huang, Gu-Quan Song, and Lin Chen.
\newblock Structural damage identification by extended k alman filter with l 1-norm regularization scheme.
\newblock {\em Structural Control and Health Monitoring}, 24(11):e1999, 2017.

\bibitem{bomarito2021development}
GF~Bomarito, TS~Townsend, KM~Stewart, KV~Esham, JM~Emery, and JD~Hochhalter.
\newblock Development of interpretable, data-driven plasticity models with symbolic regression.
\newblock {\em Computers \& Structures}, 252:106557, 2021.

\bibitem{reinbold2021robust}
Patrick~AK Reinbold, Logan~M Kageorge, Michael~F Schatz, and Roman~O Grigoriev.
\newblock Robust learning from noisy, incomplete, high-dimensional experimental data via physically constrained symbolic regression.
\newblock {\em Nature communications}, 12(1):3219, 2021.

\bibitem{sun2019data}
Sheng Sun, Runhai Ouyang, Bochao Zhang, and Tong-Yi Zhang.
\newblock Data-driven discovery of formulas by symbolic regression.
\newblock {\em MRS Bulletin}, 44(7):559--564, 2019.

\bibitem{lai2019sparse}
Zhilu Lai and Satish Nagarajaiah.
\newblock Sparse structural system identification method for nonlinear dynamic systems with hysteresis/inelastic behavior.
\newblock {\em Mechanical Systems and Signal Processing}, 117:813--842, 2019.

\bibitem{cortiella2021sparse}
Alexandre Cortiella, Kwang-Chun Park, and Alireza Doostan.
\newblock Sparse identification of nonlinear dynamical systems via reweighted ℓ1-regularized least squares.
\newblock {\em Computer Methods in Applied Mechanics and Engineering}, 376:113620, 2021.

\bibitem{rudy2017data}
Samuel~H Rudy, Steven~L Brunton, Joshua~L Proctor, and J~Nathan Kutz.
\newblock Data-driven discovery of partial differential equations.
\newblock {\em Science advances}, 3(4):e1602614, 2017.

\bibitem{schaeffer2017learning}
Hayden Schaeffer.
\newblock Learning partial differential equations via data discovery and sparse optimization.
\newblock {\em Proceedings of the Royal Society A: Mathematical, Physical and Engineering Sciences}, 473(2197):20160446, 2017.

\bibitem{long2018pde}
Zichao Long, Yiping Lu, Xianzhong Ma, and Bin Dong.
\newblock Pde-net: Learning pdes from data.
\newblock In {\em International conference on machine learning}, pages 3208--3216. PMLR, 2018.

\bibitem{long2019pde}
Zichao Long, Yiping Lu, and Bin Dong.
\newblock Pde-net 2.0: Learning pdes from data with a numeric-symbolic hybrid deep network.
\newblock {\em Journal of Computational Physics}, 399:108925, 2019.

\bibitem{both2021deepmod}
Gert-Jan Both, Subham Choudhury, Pierre Sens, and Remy Kusters.
\newblock Deepmod: Deep learning for model discovery in noisy data.
\newblock {\em Journal of Computational Physics}, 428:109985, 2021.

\bibitem{xu2021robust}
Hao Xu and Dongxiao Zhang.
\newblock Robust discovery of partial differential equations in complex situations.
\newblock {\em Physical Review Research}, 3(3):033270, 2021.

\bibitem{raviprakash2022hybrid}
Kiran Raviprakash, Biao Huang, and Vinay Prasad.
\newblock A hybrid modelling approach to model process dynamics by the discovery of a system of partial differential equations.
\newblock {\em Computers \& Chemical Engineering}, 164:107862, 2022.

\bibitem{stephany2022pde}
Robert Stephany and Christopher Earls.
\newblock Pde-read: Human-readable partial differential equation discovery using deep learning.
\newblock {\em Neural Networks}, 154:360--382, 2022.

\bibitem{rao2022discovering}
Chengping Rao, Pu~Ren, Yang Liu, and Hao Sun.
\newblock Discovering nonlinear pdes from scarce data with physics-encoded learning.
\newblock {\em arXiv preprint arXiv:2201.12354}, 2022.

\bibitem{zhang2022parsimony}
Zhiming Zhang and Yongming Liu.
\newblock Parsimony-enhanced sparse bayesian learning for robust discovery of partial differential equations.
\newblock {\em Mechanical Systems and Signal Processing}, 171:108833, 2022.

\bibitem{de1978practical}
Carl De~Boor and Carl De~Boor.
\newblock {\em A practical guide to splines}, volume~27.
\newblock springer-verlag New York, 1978.

\end{thebibliography}

\end{document}